\documentclass{article}
\usepackage{amsfonts}
\usepackage{amsmath}

\newtheorem{theorem}{Theorem}
\newtheorem{acknowledgement}[theorem]{Acknowledgement}

\input{tcilatex}

\begin{document}

\title{Rudolf Haag's legacy of Local Quantum Physics and reminiscences about
a cherished teacher and friend\\
{\small In memory of Rudolf Haag \ (1922-2016)}\\
{\small submitted to the Eur. Phys. J. H}}
\author{Bert Schroer \\
permanent address: Institut f\"{u}r Theoretische Physik\\
FU-Berlin, Arnimallee 14, 14195 Berlin, Germany}
\date{November 2016}
\maketitle

\begin{abstract}
After some personal recollectioms about Rudolf Haag and his thoughts which
led him to "Local Quantum Physics", the present work recalls his ideas about
scattering theory, the relation between local observables and localized
fields and his contributions to the physical aspects of modular operator
theory which paved the way for an intrisic understanding of quantum causal
localization in which fields "coordinatize" the local algebras.

The paper ends with the presentation of string-local fields whose
construction and use in a new renormalization theory for higher spin fields
is part of an ongoing reformulation of gauge theory in the conceptual
setting of Haag's LQP. \ \ 
\end{abstract}

\section{First encounter with Rudolf Haag}

On his return from the Niels Bohr Institute in Copenhagen to the University
of Munich Rudolf Haag passed through Hamburg to meet his colleague Harry
Lehmann, at that time the newly appointed successor of Wilhelm Lenz who held
the chair of theoretical physics since the 1920 foundation of the University
of Hamburg. It was the year 1958 shortly after the decision to construct the
DESY particle accelerator in Hamburg which created a lot of excitement. I
had nearly completed my diploma thesis under Lehmann and begun to worry
about my career.

Haag was about to accept an offer of a staff position at the University of
Illinois in Urbana, He asked me whether I would be interested to continue my
career in the US as his collaborator. The prospect of a scientific career
and the desire to change my somewhat precarious living conditions which I
encountered after my 1953 flight from East Germany to Hamburg made such a
prospect irresistible.

To better get to know each other Haag invited me to accompany him on a visit
to Daniel Kastler who at that time was a recently appointed faculty member
of the physics department of the University in Marseille. He had met Daniel
a year before and both participated in an international conference in
Lille/France where Rudolf for the first time presented his idea to base
quantum field theory on spacetime-localized operator algebras at an
international conference. Daniel was attracted by these new ideas and the
purpose of Rudolf's visit was to obtain Daniel's help for the improvement of
their at that time still shaky mathematical formulation. In this way Daniel
and Rudolf became soulmates in the exploration of what was referred to as
algebraic quantum field theory (AQFT) and later more appropriately named
"local quantum physics" which as a result of its frequent use will be
abbriviated as LQP. With Rudolf's acceptance of an offer from the University
of Illinois and his impending move to the US their collaboration was
delayed. Their first important joint publication appeared in 1962 \cite{H-K}.

The voyage to Marseille provided an opportunity to get to know each other
before my planned but not yet approved move to the US. The journey by car
through parts of Germany and across Switzerland and parts of southern France
to Marseille was an unforgettable experience. Having fled communist East
Germany and gone to Hamburg in 1953, it was my first travel outside the
German borders. In particular the journey along the C\^{o}te d'Azur with its
subtropical vegetation and its new scents and cultural impressions remaines
impressed in my memory.

After my return to Hamburg Rudolf's offer to work with him in the position
of a research associate took a concrete form; I bought boat tickets on the
Holland America line for my family and the first birthday of our daughter
was celebrated in the middle of the Atlantic.

Arriving at the University of Illinois in Urbana I encountered a formal
problem. Even taking into account the shock from the 1957 launching of the
Sputnik in 1957 which led to the creation of new positions for physicists
and engineers, the offer of a research associate position to somebody
without a Ph.D. was unusual. As I learned later from Rudolf he cleared this
problem in a conversation with the department chairman Frederick Seitz.

Frederick Seitz, a renowned physicist with politival influence on US science
policies, was a former student of Wigner. This may have played a role in
Wigner's recommendation of Haag for a full professorship at the University
of Illinois. Haag's prior visiting position at the University of Princeton
led to many scientific contacts with Wigner and Wightman. In his
reminiscenses \cite{rem} he gives credit to Wightman for having directed his
attention to Wigner's 1939 pathbreaking work on the classification of all
unitary representations of the Poincare group. It is hard to understand why
this important work of Wigner's remained unnoticed for more than a decade.

He also mentions contacts with other members of the Princeton university's
physics faculty; in particular with Valja Bargmann, who extended Wigner's
work on representation theory, as well as with Marvin Goldberger and Sam
Treiman, who at that time were working on the extension of the optical
Kramers-Kronig dispersion relations to particle physics. During this time in
Princeton Haag was the thesis adviser to Huzihiro Araki, a brilliant young
student from Japan, Araki visited Urbana several times and some discussions
even led to a joint publication \cite{AHS}.

Besides recalling personal events these notes present important ideas of
Haag's local quantum physics (LQP) in their historical context. In order to
direct attention to its largely untapped innovative strength the last two
sections include the beginnings of a LQP inspired positivity perserving
string-local renormalization theory for interactions involving higher spin $%
s\geq1$ fields whose aim to replace the "gostly" BRST gauge theory by a LQP
formulation which only uses physical degrees of freedom. For Haag this was
one of LQP's greatest challenges \cite{rem}.

Frequently occurring scientific expressions will be abbreviated: quantum
field theory (QFT). local quantum physics (LQP), point-like (pl),
string-like (sl), string-local quantum field theory (SLFT), power-counting
bound (pcb), spontaneous symmetry breaking (SSB), string theory (ST), the
Becchi-Rouet-Stora-Tyutin gauge formalism (BRST).

\section{With Haag in Urbana}

After the resounding success of renormalization theory in quantum
electrodynamics the main interest shifted to high energy nuclear
interaction. It soon became clear that these methods of perturbative
renormalization theory do not work for processes involving strong
interactions (which in the field theoretic description at that time meant
trilinear $\pi$-meson-nucleon couplings and $\pi$-selfinteractions).

Moreover doubts were increasing as to whether the locality, as formally
contained in the relativistically covariant Lagrangian quantization,
retaines its validity in the new high energy domain of nuclear interactions.
From earlier work on quantum optics it was well known that certain analytic
relations, known as the aforementioned Kramers-Kronig dispersion relations,
had a rather direct connection to relativistic causal propagation. The
problem was to derive such analytic relations for the scattering amplitudes
of strongly interacting particles.

That positivity of energy together with Einstein causality leads to analytic
properties of spacetime correlation functions (vacuum expection values) of
fields was already well known. However fields in spacetime are not directly
accessible to measurements; in experiments; one rather measures scattering
amplitudes of particles in momentum space which have a large time asymptotic
relation with fields. For those "on-shell" amplitudes in momentum space the
derivation of analytic consequences of causality posed a harder problem than
that of "off-shell" correlation functions in spacetime (Wightman functions).

For the confidence in the validity of causal locality at the higher energies
of a new generation of accelerators it was important to obtain a rigorous
derivation from the causal localization properties of field operators
(micro-causality of QFT) . The form of the expected dispersion relation was
known from the study of Feynman graphs; what was missing was a derivation
from the spacelike (anti)commutation relations of quantum fields.

The joint effort of Harry Lehmann together with Res Jost as well as
contributions from Freeman Dyson resulted in the derivation of dispersion
relation from first principles. The subsequent experimental verification at
the at that time highest energies at the Brookhaven accelerator brought the
dispersion relation project to a successful close. The confidence in the
validity of the causal locality principle in the new area of High Energy
Physics was restored and the interest in nonlocal modifications of QFT
subsided.

For Haag quantum causality is not fully accounted for by Einstein causality
in the form of (anti)commutation of operators whose spacetime localization
regions are spacelike separated. He expected that in his LQP formulation in
terms of a net of causally related algebras the quantum counterpart of
hyperbolic propagation of Cauchy data, although closely related to Einstein
causality, can not be derived from it.

The relation of LQP to Wightman's axiomatic formulation in terms of fields
and their vacuum expectation values was from Haag's LQP point of view
analogous to the relation between the coordinate-independent presentation of
modern geometry and its description in terms of coordinates. Decades later
this analogy was made more precise by H.-J. Borchers who showed that the
quantum fields form local equivalence classes and that the different members
in one class (provided they their matrixelements between the vacuum and
one-particle states do not vanish) derscibe not only the same particles and
their scattering matrix but also generate the same localized algebras \cite%
{Haag} \cite{St-Wi}.

The necessarily singular nature as operator-valued Schwartz distributions
(as a result of the omnipresence of vacuum polarization clouds) renders the
relation between fields and operator algebras very intricate. The presence
of these polarization clouds accounts for the fundamental difference of the
intrinsic causal localization and the "Born localization" in terms of an
arbitrarily chosen quantum mechanical position operator. Haag expected that
causality properties can be more natural described in his LQP setting.

In addition to Einstein causality there should also exist a time-like
causality property which is the quantum analog of the hyperbolic propagation
of classical waves. Classically the initial values within a sphere of radius 
$r$ at time $t=0$ centered at the origin have a \textit{region of influence}
which is the forward and backward light cone emanating from the sphere. But
they also lead to a compact double cone region $C=\left\{ x\
|~-x^{\mu}x_{\mu}\leq r^{2}\right\} $ inside which the radiation is
completely determined in terms of the $t=0$ Cauchy data in the sphere. Any
classical field strength measured inside $C$ which cannot be accounted for
in terms of these Cauchy data would be seen as a mysterious violation of
causal propagation since according to Einstein's causality requirement it
could not have entered from the causal complement.

Apart from free quantum fields whose propagation properties can be directly
related to those of classical fields, it is not clear how to formulate this
hyperbolic propagation property for interacting Wightman fields.
Interestingly it turns out that this is much easier in Haag's algebraic
formulation of LQP. It amounts to an equality of two different localized
(von Neumann) operator algebras 
\begin{equation}
\mathcal{A(O)=A(}\mathcal{O}^{\prime\prime})
\end{equation}
where \thinspace$\mathcal{O\ }$in the above illustration would correspond to
a in time direction slightly thickened spatial sphere\footnote{%
Haag-Kastler nets of local operator algebras are localized in open
spacetime. This is similar to interacting quantum fields which as a result
of their singular short distance behavior have to be smeared with
testfunctions supported in open regions.}. The causal complement$\ \mathcal{O%
}^{\prime}$ consists of all points which are space-like with respect to $%
\mathcal{O}$\ and the causal completion (or causal shadow) is the complement
taken twice which is generally larger than i.e. $\mathcal{O\subseteq O}%
^{\prime\prime}$ with the equality defining the maximal extension which is
consistent with Einstein causality (the causal completion).

Our joint effort was to look at the (at that time simpler appearing) problem
of the \textit{time-slice property} corresponding to the classical
determination of the future field in terms of the global $t=0\ $Cauchy data.
A global time slice can be patched together from infinitely many finite
double cone $\mathcal{O}^{\prime}s$ since the additivity property of LQP%
\footnote{%
The requirement that the operator algebra generated from the union of
localized algebras with overlapping localization regions is equal to the
algebra localized on the union of the localization regions.}\ relates a
violation of the causal completeness with that of the time-slice property.
Such a violation in a model which fulfills all other LQP requirements
implies that causal completeness \textit{is not a consequence of Einstein
causality}.

In my recollection the start of my work on "local quantum physics" (LQP)
with Rudolf is inexorably related with a beautiful summer on Wisconsin's
lake shores. Rudolf proposed to look for models which are Einstein causal
but violate causal completeness. At the beginning I was somewhat discouraged
because I considered my knowledge acquired under Lehmann as insufficient for
the new work on LQP. But it then turned out that at least some of it was
useful.

Rudolf's heuristic idea was that too many degrees of freedom within a
bounded spacetime region and a certain bound on their invariant energy
content may (a kind of relativistic phase space) lead to violations of
causal completeness. In such a case an observer would "see" more degrees of
freedom in the double cone than his experimental friend had injected into
the base region atound $t=0$. Such a "poltergeist" effect of increase of
degrees of freedom apparently coming from nowhere (since according to
Einstein causality they cannot come from space-like separated regions) is an
unexceptable violation of causality and must be excluded and the LQP setting
is the best way of formulating this.

Our simple counterexample was provided by so called generalized free fields
with a suffiently increasing mass distibution; so my modest contribution to
a joint paper consisted in some calculation with \textit{generalized free
fields} with suitable chosen continuous mass distributions $\rho(m).$
Rudolf's intuition was vindicated by the result which showed that although
all Wightman properties are satisfied, the time slice property is violated 
\cite{H-Sch}.

With the hindsight of later work one may view this illustration as a first
indication of the importance of the notion of \textit{cardinality of degrees
of freedom} which in the decades to come received various refinements; first
the Haag-Swieca 'compactness", then the Buchholz-Wichmann "nuclearity" \cite%
{Haag} and the more recent "modular nuclearity" which is used in existence
proofs of certain $d=1+1$ models of QFT (next section)

Together with the other postulates which already appeared in Haag's
contribution to the 1957 Lille conference \cite{rem} our work was a rather
complete account of the "axioms" which define the framework of LQP. Two
years later it was superseded by a paper of Haag and Kastler \cite{H-K}
which contained a more detailed account of their mathematical structure and
physical consequences. The H-K work is still considered to be the most
authoritative reference for the algebraic approach to QFT.

The time slice property played no role in most presentations of LQP but it
turned out to be important in recent formulations of QFT in curved spacetime.

For later reference it is helpful to collect the LQP causality requirements%
\footnote{%
The dash on the operator algebra refers to its commutant i.e. the subalgbre
of all bounded operators $\mathcal{A(O})^{\prime}\subset B(H)$ which commute
with $\mathcal{A(O)}$.} 
\begin{align}
& \mathcal{A(O}^{\prime})\subseteq\mathcal{A(O})^{\prime}\mathcal{\ }Einstein%
\text{ }causality,\ EC  \label{cau} \\
& \mathcal{A(O)}\mathcal{=A(O}^{\prime\prime}\mathcal{)}\text{ }causal\text{ 
}completion\text{ }property,\ CC  \nonumber \\
& \mathcal{A(O}^{\prime})=\mathcal{A(O})^{\prime},\ Haag\ duality,\ HD
\end{align}
Haag duality is an important special case of Einstein causality. The $CC$
property is closely related to $HD$. Einstein causality can be directly
expressed in terms of covariant fields, whereas for $HD$ and $CC$ this is
more subtle.

For a physical interpretation the $EC~$and $CC$ requirements are
indispensable whereas violations of $HD$ occur in the presence of massless $%
s\geq1$ interaction for multiply connected spacetime regions (tori). The
most prominent illustration is the operator of magnetic flux through a solid
torus $H(\mathcal{T}\mathfrak{)}\mathcal{\ }$which belongs $\mathcal{A(T}%
^{\prime })^{\prime}$ but not to $\mathcal{A(T})$ \cite{LRT}.
Interestingly~the Aharonov-Bohm effect is related to this breakdown of Haag
duality \cite{beyond}.\ 

The causal completeness property also severely limits relations between LQPs
in different spacetime dimensions ("extra dimensions"). This affects in
particular the mathematical isomorphism between LQPs on $n$\ dimensional
Anti de Sitter space (AdS) and $n-1$ dimensional conformal spactime which
share the same spacetime symmetry group. On heuristic grounds one expects
that the AdS-CFT isomorphism leads to similar problems as the previous $CC\ $%
violation for certain generalized free fields with too many degrees of
freedom.

This is precisely what happens on the lower spacetime dimensional conformal
side \cite{Re}. In fact starting with a $CC$ obeying AdS free field \cite%
{Du-Re} one obtains an "overpopulated" conformally invariant generalized
free field \cite{H-Sch}, and this problem does not disappear in the presence
of interactions. Conversely the LQP double cone algebras on the $AdS$ side
obtained from a "healthy" conformal LQP are "anemic" in the sense that
compact localized algebras do not contain any degrees of freedom and one has
to pass to noncompact localization regions to encounter algebras which are
not multiples of the identity. All those cases the algebraic isomorphism
preserves $EC$ but violates the with degrees of freedom connected $CC.$

The Kaluza-Klein proposal of extra or lowered spacetime dimensions works for
classical field theories as well as semi-classical approximations but it
clashes with QFT$.\ $ "Transplanting" the matter content between worlds of
different spacetime dimensions preserves $EC$ but fails on $CC$.

The issue of causal localization sustaining quantum degrees of freedom is a
very subtle one which is inexorably related with the role of vacuum
polarization clouds in causal localization and has no counterpart in quantum
mechanics or classical field theory. Using the standard formulation of QFT
field coordinatizations one may easily overlook the breakdown of the causal
completeness property as a result of an overpopulation of degrees of freedom
resulting from resettling the degrees of freedom of a higher dimensional LQP
into a lower dimensional spacetime vessel. This is precisely what happened
when the AdS-CFT isomorphism and the idea of extra dimensions became a focal
point of interests in the 90s which led to thousands of publications.

During the almost 3 years of my time in Urbana there were many interesting
visitors. I remember that Gell-Mann on one of his visits asked us if we had
a more intrinsic understanding of the relation between the partially
conserved axial vector currents (PCAC) with the field of the $\pi$-meson. At
that time gauge theoretic Lagrangian models with axial $\rho$-mesons as
proposed by Sakurai enjoyed great popularity. At the end of the discussion
Murray Gell-Mann joked: "you mean we can shoot Sakurai?" before he enjoyed
looking at our somewhat helpless expressions.

Together with Haag I participated in a summer school in in Boulder,
Colorado. My remembrances about the activities in physics are faint but I do
recall having been impressed by the beautiful nature of the Rocky Mountains
and a subsequent journey with my family through the Yellowstone National
Park.

I also recollect an extremely peculiar occurrence. When I looked as usual
into the weekly Time Magazine I came across a story about two mathematicians
at the University of Illinois which were engaged in classified work for the
NSA before they defected via Cuba to the Soviet Union taking classified
material with them. The name of one of them was the same as that of somebody
who lived in an apartment in Urbana which I rented shortly before I went to
Boulder. The apartment in a university housing project became too small for
my family after the birth of my son. The former tenant whose name was Martin
also sold his piano and some furniture to me before he moved out. There was
a picture of the two mathematicians in Time Magazine, but the quality of
printed photos in those times was so poor that I could not identify him. I
brushed the incidence aside as a coincidence of names and enjoyed the rest
of the stay.

Two agents of the CIA were already waiting for me. Apparently they found the
check of my payment for the piano in a Washington deposit. They really knew
a lot about my past, in particular that in 1953 I fled from East Germany.
Probably they obtained their knowledge from an archived protocolled hearing
in a transit refugee camp, a former concentration camp near Hamburg where
besides German officials also a US officer was present.

Rudolf assured me that this matter will be cleared up in a short time.
Indeed after several meetings in a restaurant I succeeded to convince them
that my involvement was coincidental and that I was not an East German spy.

Many years later when I mentioned the Martin-Mitchell spy story at an
international physics conference to Ludvig Faddeev, he told me that a week
before both of them applied for a position at the Steklov Institute in
Leningrad. By that time they had Russian wifes and and families. How was
this possible; did the communist ideology convert two homosexuals ?

Before my position at the University of Illinois came to an end, I met Jorge
Andre Swieca who, after having spent a year at the Werner Heisenberg
Institute in Munich (one of the largest Max Planck Institutes for physics in
Germany), passed through Urbana on his way to Brazil. The purpose of this
visit was to introduce himself to Rudolf as his new Research Associate.
After he defended his Ph.D thesis at the University of \ Sao Paulo (with
Guettinger as his advisor) he returned to Urbana to start his work with Haag.

During my stay in Urbana I had obtained some results which were appropriate
to be used for a Ph.D thesis. I returned 1963 to Hamburg where I submitted
my thesis. The terminology "Infraparticles" in its title \cite{FdP} referred
to the conjecture that the infrared divergencies which appear in the
scattering amplitudes of electrically charged particles are related to a
modification of the Wigner particle structure. I was able to illustrate this
in a two-dimensional model. The realistic case was taken up two decades
later by Buchholz. The issue of infraparticles has remained a challenging
topic of LQP \cite{Haag}

After a one year at the IAS in Princeton, a short stay at the University of
Hamburg and a visit to the Middle East University in Ankara at the
invitation of Feza Gursey I returned to the US to take up my new position of
associate professor at the University of Pittsburgh.

Shortly before I left Urbana I shared an office with Derek Robinson who
became Haag's second collaborator. During a visit by Kastler, Robinson
Swieca and Kastler investigated the properties of conserved currents within
the new algebraic Haag-Kastler setting of LQP. They found that the
conservation law only secures the existence of a "partial" charge which
secures the existence of a local symmetry within each finite spacetime
localization regions but that the global charge may diverge i.e. the inverse
of the quantum Noether theorem may be violated i.e. the current conservation
may not secure the existence of a global "charge" (the infinitesimal
generator of a unitary symmetry.

It was known from perturbative investigations of self-interacting scalar
fields by Goldstone that the local current conservation may lead to a
divergent global charge resulting from the contribution of a massless scalar
("Goldstone") boson which impedes the large distance convergence and in this
way causes a situation which was appropriately referred to as spontaneous
symmetry breaking (SSB).

Kastler, Swieca and Robinson showed that this cannot happen in the presence
of a mass gap \cite{HKS}, and in a follow up paper (based on the use of the
Jost-Lehmann-Dyson representation) Swieca together with Ezawa \cite{E-S}
succeeded to prove the Goldstone theorem in a model- and perturbation-
independent way\footnote{%
The Goldstone theorem states that a N\"{o}ther symmetry in QFT is
spontaneously broken precisely if a massless scalar "Goldstone boson"
prevents the convergence of some of the global charge $Q=\int j_{0}=\infty.. 
$}. Goldstone constructed renormalizable SSB models of self-interacting
scalar particles by applying the "shift in field space" prescription to
formally symmetry-preserving "Mexican hat potentials".

This quasiclassical prescription leads to a model-defining first order
interaction density which maintains the conservation of the symmetry
currents in all orders. There are symmetry-representing unitary operators
for each finite spacetime region $\mathcal{O}$ but the global charges $%
Q=\int j_{0}$ of same symmetry generating currents diverge. This is the
definition of SSB whereas the shift in field space procedure is a way to
prepare such a situation whenever SSB is possible.

For the later presentation of the Higgs model it is important to be aware of
a fine point about SSB whose nonobservance led to a still lingering
confusion. As soon as scalar self-interacting fields are coupled to $s=1$
potentials the physical interpretation of the field shift manipulation on a
Mexican hat potential as a SSB is incorrect; one obtains the Higgs model for
the wrong physical reasons and misses the correct reasons why there can be
no self-interacting massive vectormesons without the presence of a $H$%
-field. Although this can be decribed correctly in the gauge theoretic
formulation, a better understanding is obtained in the positivity preserving
string-local setting of LQP (see section 6)

QFT is not a theory which "creates" masses of model-defining fields. The
masses of those free fields which define the first order interaction
density) are, together with the coupling strengths, free parameters\footnote{%
Masses and mass ratios may appear in coupling strengths of induced higher
order contributions.}. The only "dynamic" masses are those of bound states
created by acting with interacting composite fields on the vacuum state but
unfortunately there is no perturbative methods which descibes bound states.
space.

In a later paper Haag and Swieca investigatigated \ the cardinality of
states contained in a finite spacetime region with limited energy content 
\cite{H-S}. In quantum mechanics this corresponds to the number of degrees
of freedom per cell in phase space which is finite. They found that LQP
leads to an infinite set whose cardinality cannot exceed that of a compact
set.

\section{The Brazilian connection}

Shortly before I left in 1962 I met Jorge Andre Swieca for the first time
when, on his return from the Max Planck Institute in Munich to the
University of Sao Paulo (USP) he passed through Urbana for an interview with
Haag. His thesis adviser was Werner G\"{u}ttinger who, as several other
German physicists, was invited to the in 1952 newly founded Instituto de
Fisica The\'{o}rica (IFT) in Sao Paulo. G\"{u}ttinger recognized the
potential of Andre Swieca and arranged a visiting position for him at the
MPI in Munich. When I met Andre in Urbana he was on his way back to the USP
in order to defend his thesis before taking up the research associate
position with Haag in Urbana.

G\"{u}ttinger is one of the few theoretical physicists who, shortly after
Laurant Schwartz's presentation of the theory of distributions, saw the
relevance of that theory for the description of the singular nature of
quantum fields. Before he obtained a permament position at the University of
T\"{u}bingen/Germany he spend some years in the second half of the 50s at
the ITF. It is interesting to note that around 1952 Laurant Schwartz
together with Alexander Grothendiek spend some time at the USP. On my first
visit in 1968 there still existed traces of the legacy of Laurant Schwartz
in the form of courses on distribution theory at the USP physics department
which were presented by a young Brazilian lady who obtained her PhD with
Laurant Schwartz.

My first chance to take a short leave of absence from the University of
Pitteburgh to follow Andre Swieca's invitation to the USP came in 1968.
After his return from the collaboration with Haag in Urbana to Brazil at the
end of 1966 Andre held the position of a junior professor at the USP. When I
arrived he was surrounded by a group of enthusiastic young students of whom
the most advanced (Jose Fernando Perez) was assigned the task to take care
of me and to help me with the written version of my lectures on QFT. This
was the beginning of what Haag in his reminiscences called the "Brazilian
connection" (\cite{rem} page 24).

During my visit Andre received the Moinho Santista prize for his quantum
field theoretic work on symmetries and their spontaneous breaking. After
Jaime Tiomno, one of the founders (together with Mario Schemberg and Jose
Leite-Lopes) of theoretical physics in Brazil, Andre was its second
recipient. After his collaboration wth Kastler and Robinson in Urbana on the
LQP formulation of symmetries and their conserved currents he had pursued
this issue in more depth with particular attention for spontaneous symmetry
breaking (SSB) for which the current remained conserved but the presence of
a massless Goldstone boson one looses the symmetry generator since the
global charge diverges. In a joint paper with H. Ezawa the Goldstone
theorem, which previously only existed as a perturbative property of a
special class of models, was derived in a model-independent way from the
causal localization principles of QFT \cite{E-S}.

He lectured on his results in Erice \cite{Er} and up to date I know no
clearer model-independent presentation of Goldstone's SSB theorem about SSB
as a consequence of the causality and spectral properties principle of QFT
than that in his notes. This is particularly important in times in which SSB
became somewhat misleadingly synonymous with a shift in field space on a
Mexican hat potential (see remarks in previous section).

At the time of my visits during the 60s and 70s Brazil was ruled by a
military junta whch took power in 1964 coupe. At the start of my visit in
1968 I hardly noticed the presence of a military dictatorship, but the
situation changed abruptly in May 1968 when at the time of intensification
of the Vietnam war there were student demonstrations in Paris and Berlin and
other places. I listened to the news on my short wave radio but soon became
aware that there was an increasing number of demonstrations against the
military dictatorship whose only connection with the Vietnam war was that
those who started the war were the same who supported the military regime in
Brazil.

Many years after Swieca's premature death in 1980 somebody asked me whether
I knew something about a rumor that after having received the Santista prize
he was approached by the military government to explore the possibility to
offer the post of a scientific/cultural attache to Israel. I did not, but I
was sure that if this really happened Andre would have declined an offer of
representing a military dictatoship in a democratic country. Sometimes I saw
military police entering the USP campus and later I learned that one of my
collegues Ernesto Hamburger was taken into custody and his wive was tortured.

Andre told me a saddening story about an occurrance which happened shortly
before to one his professors from whom he took his first physics courses.
Plinio Susskind had a very strong personal contact with his students, after
the lectures he joint them to continue discussions about matters of physics
and daily events in nearby cafes and bars. He had a collection of books
which included the work of Marx and others which after the military coup
were considered subversive. When the military police searched his apartment
they found a copy of Sergei Eisenstein's "Couracado Potemkin" (Battleship
Potemkin). He was taken into custody and after having been released he lost
his university position.

He was not internationally known and had no chance to continue working
outside Brazil. He fell into poverty and Andre and some of his former fellow
students supported him for many years. The worst aspect of a military regime
is that it encouraged denunciations which in some people used to settle
accounts. Two of the founders of theoretical physics in Brazil as Jaime
Tiomno and Leite-Lopes who felt threatened by the regime accepted positions
in the US or France. For more than a decade, starting from the beginning of
the 70s up to the return of democracy in 1985, the catholic university of
Rio de Janeiro (PUC) became a refuge for many Brazilian scientists including
Jorge Andre Swieca who worked there for several years.

On this first visit to Brazil there was little time and peace of mind to
talk about how to use our shared knowledge acquired as collaborators of Haag
for establishing a joint project. We postponed the discussions of topics of
joint interests to future visits.

One week after my return to Pittsburgh I received a notice that military
tanks entered the USP at dawn and took positions around the CRUSP housing
and took everybody into custody. Apparently was released on the same day;
not because he was particularly cooperative but rather as the result of
taking notice that his fiance was the daughter of a high ranking military;
an occurrence which is easily understood for those who experienced the
Brazilian "jeitinho" which survived any system up to date.

When back in Pittsburgh I obtained informations about the worsening
political situation in Brazil I found myself in a unusual schizophrenic
situation; here I was living peacefully in a democratic country whose
government supported military dictatorships in other countries under which
my colleagues suffered.

Less than two years later Andre visited me at the University of Pittsburgh
where the QFT group was meanwhile strengthened by Ruedi Seiler, a
mathematical physicist who received his PhD shortly before from the ETH in
Zurich. Looking for a topic on which one could start a short time
collaboration we found it worthwhile to investigate to what extend Einstein
causality and the causal shadow property retain their validity for
interactions of quantum fields with external (classical) fields.

Using functional analytic methods it was possible to show with the help of
the energy norm that these causality properties hold for models of low spin
quantum fields coupled to time-dependent asymptotically vanishing classical
fields and for $s>1$ interactions we extended previous observations about
acausalities \cite{SSS}. In case of strong stationary external fields we
were able to improve the understanding about an inconsisteny of the Klein
Gordon field in a strong potential made thirty years before \cite{SS}. The
result was that there are two ways of quantizing bound states with negative $%
E^{2}~$namely either by using indefinite metric or by abandoning the vacuum
postulate and accepting repulsive (inverted) oscillator degrees of freedom
associated to such bound states..

This led me to take another look at "tachyons" described by fields with $%
m^{2}\rightarrow-m^{2}.$ As the name suggests these fields were thought of
as being associated to fields describing "superluminal stuff". But how can
this be in view of the fact that a classical tachyon field has a perfectly
causal propagation? The answer was that in limiting oneself to real
spacelike momenta one has left out imaginary values $-m^{2}+\vec{p}^{2}<0$
whose momenta lead to inverted oscillators which in quantum theory requires
to substitute the vacuum state by a continuum of negative energy "jelly"
states. Such a situation without bottom becomes chaotically unstable in the
presence of interactions; this is reminiscent of Dirac's hole theory except
that in the tachyon case there is no "filling". with arbitray large negative
energies They correspond to those inverted oscillators which in the problem
of strong external potentials prevent the existence of a lowest energy
vacuum state. In the tachyon problem they require the introduction of a
continuum of negative energy "jelly states" whose presence is indispensable
for maintaining causal propagation. Although the free theory exists, any
perturbation will cause a similar instability as the Dirac sea before
filling it, except that for tachyons such filling is not possible \cite%
{tachyon}.

This instability argument was later used in the quasiclassical preparation
of spontaneous symmetry breaking (SSB), which is a mild form of symmetry
breaking in which there still exists a conserved current but its charge (the
generator of the symmetry) diverges due to the presence of a massless scalar
boson (the Goldstone boson). Since it is somewhat tedious to prepare a first
order interaction density with this property one starts from a symmetric
Mexican hat kind of selfinteraction \ of a multiplett of particles and uses
the quasiclassical trick of a shift in field space which brings an
apparently tachyonic situation of a Mexican hat potential into a less
symmetric one with a vacuum.

The test whether the quasiclassical shift in field space on a
selfinteracting multiplet with a tachyonic mass term which preserves the
current conservation of the multiplett leads really to a SSB is decided in
terms of $Q=\infty .~$This and not the manipulation is the definition of
SSB. In the absence of couplings to $s\geq1$ fields this is the case but it
fails in models in which the scalar matter couples to a vector potential. As
will be demonstrated in the last two sections the "fattening of the photon"
does not require the presence of a Higgs field; it is rather related to the
appearance of an escort field which in turn is the unavoidable consequence
of maintaning positivity in the presence of massless vector potential.

As far as one knows Nature provides no realization of exact internal
symmetries or SSB in particle physics beyond the particle-antiparticle
symmetry; the application remains in the hands of phenomenologists. But
there can be no doubt that Nature supports the existence of a Higgs particle
without which there can be no self-interacting massive vector mesons.

Shortly after my 1968 visit of the USP John Lowenstein, Haag's last PhD
student before he left Urbana and moved to the University of Hamburg, joined
Andre as a post-doc. Their joint work on QED in two dimensions \cite{L-Sw}
impressed me as a thorough application of mathematical ideas and concepts
from LQP. For private reasons John wanted to return to the US and I was able
to be helpful in obtaining a position at the University of Pittsburgh. This
was the start of a fruitful collaboration on perturbative renormalization
theory to all orders, in particular to gauge theory and its axial anomalies,
which continued after my move to Berlin in 1971.

The collaboration with Andre and his research group continued during the
70ies; he came twice to Berlin and I met him 3 times, twice at the PUC in
Rio and a third time after he moved to the USP in Sao Carlos. We wrote
several papers on models with conformal symmetry in particular on global
operator expansions and one of my collaborators (A. Voelkel) had a short
time visiting position at the PUC.

In the middle 70ies a class of two-dimensional models with factorizing
S-matrices became the focus of attention. These integrable models were of
particular interest since perturbative constructions did not permit to
establish the existence of nontrivial models so that QFT was the only area
of theoretical physics for which the existence of interacting models within
its conceptual framework (causality, Hilbert space positivity) remained
widely open.

The discovery of these $d=1+1$ integrable models led to a close
collaboration between group of research associates at the FU Berlin (Berg,
Karowski, Thun, Truong and Weisz) with a group around Swieca (K\"{o}berle,
Kurak, Marino, Rothe) with myself representing the link between the two.
Swieca' death at the end of 1980 also marks the end of what Haag in his
reminiscences called "the Brazilian connection".

A collection of Swieca's publications appeared later as "obras colligidas" 
\cite{obras} to which I wrote a long introduction with the title "From the
Principles of Quantum Field Theory towards New Dynamical Intuition from
Studies of Models". This marked the end of a decade lasting collaboration to
explore and illustrate the content of Haag's LQP in concrete models of QFT.
Our last joint project to detach operator product expansion from conformal
QFT and in this way obtain a nonperturbative construction remained an
unfulfilled project. Recently there has been significant progress on this
old problem \cite{H-H}.

When, I revisited Brazil 20 years later, some of Andre's closest colleagues
had retired or died (Jose Giambiagi) and his former younger collaborators
were working on different problems.

The old project received a new impulse when Karowski and Weiss extended it
to what nowadays is referred to as the "formfactor program" which consists
in the explicit nonperturbative construction of matrixelements of fields
between particle states. Besides presenting new insights into
nonperturbative QFT its aim is to construct a QFT in terms of vacuum
expectation values of quantum fields which can be formally represented as
infinite sums over product of formfactors. As in perturbation theory there
is presently no control of such sums.

Meanwhile a different approach has led to the first existence proofs for
integrable models in the absence of bound states. It does not use individual
fields but rather directly Haag's LQP setting in terms of net of local
algebras. It is a top-to-bottom approach which starts from the observation
that the modular localization theory (see next section) connects the
algebraic structure of wedge-localized algebras with the S-matrix and uses
the fact that for factorizing S-matrices without bound states there exist
simple generating operators for wedge algebras whose Fourier transforms
fulfill the Zamolodchikov-Faddeev algebra relations.

Knowing the structure of the wedge algebra the next step is to show the
existence of nontrivial algebras associated to compact spacetime regions
resulting from intersection of wedges. This is the real hard part where
estimates about degrees of freedom in the form of "nuclear modularity" enter 
\cite{Al-Le}. The terminology top-to-bottom refers to obtain algebras of
compact spacetime regions by intersection of wedge algebras. Covariant
fields which generate these algebras would appear only in a later stage of
this ("top-to-bottom") construction. Since the physical consequences can be
directly extracted from the algebras they are not needed. The protagonists
of these ideas belive that future existence proofs of interacting QFTs in $%
d=1+3$ will be based on such top-to-bottom constructions.

The remainder contains some remarks which bear no relation to physics but
which form part of my personal "Brazilian connection"

During the collaboration with Andre the weight of the past was always
present. Andre was born in 1938 in Warsaw/Poland. His family had the good
luck to escape from the murderous anti-semitism of the Nazis to that part of
Poland which in 1939 according to the Hitler-Stalin pact was occupied by the
Soviet from Union. Before Hitler's assault of the Soviet Union and the Nazi
occupation of the rest of Poland, the Swiecas fled to the Soviet Union from
where they succeeded to reach Vladivostoc on the transsiberian railroad from
there they got by boat to Yokohama and finally to South America.

They had some relatives in Rio de Janeiro but Getulio Vargas's anti-semitic
police chief Filinto M\"{u}ller created problems which forced them to remain
for some months in Buenos Aires. In the 70s M\"{u}ller was the senator of
the states of Mato Grosso and leader of the Arena party which was created by
the military.

I was invited several times to the house of Andre's parents and on one of
these visits I sensed a mood of commotion. It was the day on which Filinto M%
\"{u}ller died in a plain from Rio to Paris. In those days the seats in many
airplanes contained a material (polyvinyl chloride) which, if ignited by a
cigarette, could lead to a smoldering fire. This happened on M\"{u}ller's
flight; the captain made an emergency landing but all the passengers and
those of the crew who did not succeed to enter the captains cabin perished
in the toxic fumes.

Andre and his parents were not religious, yet there was a feeling of higher
form of justice since Filinto M\"{u}ller was responsible for the deportation
of Olga Benario-Prestes on a Spanish ship via Franco's Spain and her
extradition to Nazi-Germany \cite{book}. Olga, a German communist, together
with the Brazilian tenent Luis Carlos Prestes were in opposition to the
dictatorship of Getulio Vargas. Their attempt to initiate a revolt within
the Brazilian military failed and both were jailed. M\"{u}ller deported Olga
om a Spanish ship and the Franco extradited her to Nazi-Germany. Being of
jewish descent this was like a death penalty. Her deportation caused
national and international protests in particular since such an extradition
in a state of advanced pregnancy was against the Brazilian law. Olga gave
birth to her daughter Anita Leocadia Prestes in a Berlin prison clinic.
Using her connections to the Itamaraty (the Brazilian Foreign Office) \
Prestes mother succeeded to take the baby to Brazil. Nowadays she is a
professor of history at the Federal University of Rio de Janeiro.

Olga was taken to the Ravensbr\"{u}ck concentration camp and killed by gas
in the Bernburg Euthanasia Centre which the Nazis created years before as
part of their euthanasia program for mentally ill people. When after
protests from part of the catholic church this clandestine murderous progran
was halted, the installation was used to kill prisoners from those near by
concentration camps which, as the women's camp at Ravensbr\"{u}ck, had no
extermination facilities.

The fate of Filinto M\"{u}ller, who died the same way as Olga Benario, is
remarkable even for those who do not believe in higher justice and destiny.

The fact that I spend my childhood in Bernburg, and that I remembered my
mother wispering with neighbors about busses with painted windows arriving
at the mental hospital, constitutes an encouter with my past in a manner
which I could never have imagined. In this way this became an inexorable
part of my "Brazilian connection".\ 

\section{Local Quantum Physics and Modular Localization}

One of the meetings between mathematical physicists and mathematicians which
I attended in the 60s was a 1967 conference in Baton Rouge, Luisiana. In the
center of attention was the work of Yuji Tomita, an elderly Japanese
mathematician who appeared with a thick still unpublished manuscript. Its
title contained the word "modular", indicating that he wanted his new
results on operator algebras to be viewed as a kind of noncommutative
generalization of measure theory. I understood very little of these new
mathematical results.

Richard Kadison, an authority on operator algebras who chaired the
conference, had doubts about some of Tomita's arguments. He encouraged
Tomita's compatriot Takesaki to review the arguments and rework the
presentation of the results together with Tomita. This led to the still
authoritative first book on Tomita's theory which became known as \textit{%
the Tomita-Takesaki modular theory}$\ $\cite{Ta}.

Tomita's ideas led to a new formulation in which the unitary \textit{modular
group} was associated with operator algebras satisfying certain conditions.
This seemed to be connected in some way to new formulation (statistical
mechanics of open systems) of equilibrium statistical mechanics directly in
the infinite volume limit (without using Gibbs trace formula) proposed by
Haag, Hugenholtz and Winnink \cite{H-H-W}. Their results were also presented
at this conference. What these authors referred to as the KMS property had
its much more general operator-algebraic counterpart in Tomita's work.\ 

The KMS\ property appeared first in previous work by Kubo, Martin and
Schwinger (the historic origin of the terminology). In the work of these
authors it was merely a computational trick which converted the calculation
of traces in the Gibbs formula into more manageable analytic properties
involving analytic continuations. But in the new context it acquired a
foundational meaning far beyond a mere computational device.

This terminoly and also some of its physical content was afterwards adopted
by the operator algebraists; Alain Connes and Uffe Haagerup used it in his
impressive classification of the type III von Neumann factor algebras.
Whereas the mathematical concepts of quantum mechanics, such as the Hilbert
space and operators acting on it, existed before its discovery, the
foundations of modular theory are the result of a joint effort between
mathematicians and mathematical physicists. More details about the
path-breaking Baton Rouge conference can be found in Haag's reminiscences 
\cite{rem} and an authoritative account about its impact on mathematical
physics including an important interrelation with causal localization is
contained in a seminal article by H.-J. \ Borchers \cite{Bor}.

A decade later the work of Bisognano and Wichmann \cite{B-W} revealed that
causal localization and the KMS thermal aspects are inexorably
interconnected; subsequently Geoffrey Sewell pointed out that this
interrelation plays a fundamental role in the understanding of Hawking's
black hole radiation and the Unruh effect \cite{Sew}. An account of the
Hawking radiation from the viewpoint of LQP can be found in \cite{Fr-Ha}.

The interest in the application of LQP to problems in curved spacetime is
reflected in an increasing number of publications starting in the 90s. A
recent review with references to earlier work can be found in \cite{Fr-Rej}.

In this context it is also interesting to note that modular localization
sheds some new light on a fascinating but for a long time incompletely
undertood controversy between Einstein and Jordan (the Einstein-Jordan
"conundrum") which led Jordan to the first model of a field theory (the
model of conserved current in $d=1+1$).

Haag's view of localized quantum matter as a net of causally localized
operator algebras acting in a joint Hilbert space received important support
from the modular theory of operator algebras \cite{Fr}. A particularly
fruitful conceptual enrichment came from the application of \textit{modular
localization }to integrable models\footnote{%
For a recent account containing a rather complete list of references to
previous work see \cite{Al-Le}.} and to the use of Wigner's representation
theory of the Poincare group for the construction of noninteracting nets of
operator algebras \cite{BGL}\textit{.}$\ $Since this concept plays an
important role in the later sections an at least rudimentary understanding
will be helpful.

There exists a weaker version of the T-T modular theory which does not refer
to operator algebras but uses the concept of a so-called \textit{standard
subspace }$\mathcal{K}$ of a Hilbert space $\mathcal{H}$. This is a closed
real subspace $\mathcal{K\subset H}$ whose complexification is dense $\ $in $%
\mathcal{H}\ $i.e. $\overline{\mathcal{K+}i\mathcal{K}}\mathcal{=H\ }$and $%
\mathcal{K\cap}i\mathcal{K}=\left\{ 0\right\} \ $where the bar referes to
the closure.$.$ The Tomita $S$ operator is then defined as $S(\zeta
+i\eta)=\zeta-i\eta$; conversely $\mathcal{K}$ can be represented in terms
of a Tomita operator $\mathcal{K}=\ker(S-1).$ As in the algebraic
Tomita-Takesaki setting the $S$ has a polar decomposition $S=J\Delta^{1/2}$
where $\Delta ^{it}$ is an automorphism of $\mathcal{K}$ and the antiunitary 
$J$ transforms $\mathcal{K}$ into its symplectic complement (symplectic
orthogonal) of $\mathcal{K}\ $within $\mathcal{H}$ which is defined in terms
of the symplectic product $i\func{Im}(f,g).$

The simple physical illustration of the connection of causal localization
with the T-T modular theory is provided by the 2-point function of a scalar
free field%
\begin{equation}
\left( f,g\right) =\left( \varphi(f)\Omega,\varphi(g)\Omega\right)
,~\varphi(f)=\int\varphi(x)f(x)d^{4}x   \label{scalar}
\end{equation}
defines a scalar product between the forward mass-shell restriction of the
Schwartz test functions. The Hilbert space $\mathcal{H}$ is the space of
Wigner wave functions $\mathcal{H}$ of a scalar particle which is obtained
from the closure of the forward mass-shell restriction of the Fourier
transformed test functions.

The closed subspace $\mathcal{K(O})$ of $\mathcal{H}\ $obtained by the
closure of real test functions localized in $\mathcal{O}$ turns out to be
standard in the above sense; this follows from the one particle projection
of the cyclic and the separating property of quantum fields known as the
Reeh-Schlieder property \cite{Haag} (or can be shown directly). Since $i%
\func{Im}(f,g)$ can be written in terms of the vacuum expectation of the
commutator%
\[
i\func{Im}(f,g):=\left( \Omega\left[ \varphi(f)^{\ast},\varphi(g)\right]
\Omega\right) 
\]
the aforementioned symplectic orthogonality receives a physical
interpretation in terms of Einstein causality.

More interesting and important is the inversion of this relation i.e. the
construction of a net of causally localized subspaces $\mathcal{K(O})\subset%
\mathcal{H}_{Wig}$ of the Wigner's representation space using his
representation theory of the Poincare group. The key for this construction
is the \textit{Bisognano-Wichmann property} i.e. the physical identification
of the Tomita operator $S_{W}$ for a wedge region e.g. $W=\left\{
x_{3}>\left\vert x_{0}\right\vert \right\} $ $\ $In this situation these
authors showed (under mild technical assumptions in a Wightman setting \cite%
{B-W}) that the antilinear $S$-operator associated to the dense set of
states obtained by applying a wedge-localized operator algebra to the vacuum
can be expressed in terms of phy,sical data. Whereas the unitary modular
group $\Delta^{it}$ associated to the radial part of its polar decomposition 
$S=J\Delta^{1/2}$ is the $W$-preserving boost $\Delta^{it}=U(\Lambda_{W}(-2%
\pi t))),$ the antiunitary angular part $J$ is, apart from a $\pi$-rotation
in the plane of the edge, the TCP operator which plays a funde,ental role in
QFT.

With this physical identification one obtains the modular subspace $\mathcal{%
K}(W),$ and by covariance also all its Poincare transforms. Subspaces $%
\mathcal{K(O})$~for general localization regions $\mathcal{O}$ are obtained
in terms of intersections, their modular groups have no geometric
interpretation (except in the presence of conformal covariance); although
they preserve the localization region their action inside is "fuzzy" i.e.
cannot be visualized in terms of geometric transformations inside $\mathcal{O%
}$. \ Using the functorial relation between real one-particle subspaces and
operator subalgebras, which is defined in terms of Weyl operators, one
finally arrives at an explicit construction of Haag's net of local algebras
in the absence of interactions \cite{BGL}.

This functorial relation which maps localized real subspaces into local von
Neumann algebras in such a way that subregions correspond to subalgebras and
Einstein causality holds permits a generalization to all positive energy
Wigner representations. The fact that it is tied to the energy positivity
shows s (perhaps somewhat unexpected) close connection between geometric
with spectral properties of the translation operators (stability properties).

This relation breaks down in the presence of interactions. In this case one
may start from the Wigner-Fock space which in the presence of a gap is
provided by (LSZ or Haag Ruelle \cite{Haag}) scattering theory. It has been
known for a long time that in this case the TCP operator differs from that
of a free incoming fields by the scattering matrix $S_{scat}$ so that one
obtains $J=S_{scat}J_{0}$ where $J_{0}$ refers to the free fields associated
to the Wigner-Fock space.

An explicit construction of modular localized subspace in the presence of
interactions is possible for integrable models with known factorizing
S-matrix. An important supporting property is the existence of so called 
\textit{vacuum polarization free generators} (PFG) i.e. operators in an
interacting theory whose application to the vacuum creates a
polarization-free one-particle vector. Their existence is based on the
relation between the tightness of causal localization and the strength of
interaction-caused vacuum polarization clouds. It is well-known that the
singular nature of states created by interacting quantum fields is related
to the strength of vacuum polarization clouds; the larger the spacetime
localization region conceded to the clouds, the easier to find less singular
operators.

It had been known for a long time that covariant point-local fields which
create a polarization-free one particle state from the vacuum must be free
fields (the J-S theorem \cite{St-Wi}). The more general concept of vacuum
polarization free generators (PFG) leads to theorem that for compact
localized spacetime regions such PFGs do not exist in interating theories.
The tightest noncompact region for which (under certain weak condition) PFGs
exist are (arbitrarily narrow) space-like cones \cite{J-S}. The fact that
they are always available in wedge regions \cite{BBS} makes the latter an
ideal point of departure for existence proofs.

In the case of integrable models such PFGs are provided by the Fourier
transforms of the creation/annihilation operators which obey the
Zamolodchikov-Faddeev algebra \cite{AOP} whose commutation structure is
given in terms of the known elastic part of the factorizing S-matrix. This
observation is the starting point of an LQP-based construction project
starting from the PFG-generated $\mathcal{A}(W)$ operator algebra. A highly
intricate part of the construction is the demonstration of nontriviality of
double cone intersections of wedge algebras \cite{Al-Le}. Here concepts of
cardinality of degrees of freedom in the form of \textit{modular nuclearity}
play an important role.

The obtained results are complementary to those of the form factor program
for integrable models . Whereas the latter lead to concrete closed-form
expressions for formfactors of point-local fields (but without control of
the convergence of the resulting infinite series expressions for the
correlation functions), the LQP construction starts from the generators of
the wedge algebra and establishes the existence of a nontrivial double cone
intersection of arbitrary small diameter (but falls short of constructing
the generating point-local fields).

Interacting models in dimensions $d>1+1\ $are not integrable and hence
possess no closed form (analytically representable) solutions. Whether the
extension of ideas based on modular localization to $d=1+3$ dimensional
interacting models will lead to a nonperturbative control remains a dream of
the future. Different from all other areas of theoretical physics QFT
remains an enigmatic project.

QFT earned its standing as the most comprehensive description of Nature's
physical properties from the observational success of its perturbative
formulation. The predictive success of the Standard Model is based on low
order perturbation theory complemented by phenomenologically supported
proposals.

Contrary to a widespread misconception renormalized perturbation theory does
not depend on any quantization parallelism with classical field theory. As
shown in \cite{Wein} covariant point-local (pl) free fields are constructed
from Wigner's theory of unitary positive energy representations of the
Poincar\'{e} group; the corresponding spaces of particle wave functions bear
no relation to actions of classical point-local particles (section5).

In terms of the creation/annihilation operators $a^{\#}(p,s_{3})$ for
massive particles and their anti-particles $b^{\#}(p.s_{3})$ which act in a
Wigner Fock space, the pl covariant free fields are of the form%
\begin{equation}
\psi^{A,\dot{B}}(x)=\frac{1}{\left( 2\pi\right) ^{3/2}}\int\dsum
\limits_{s_{3}=-s}^{s}(e^{ipx}u^{A,\dot{B}}(p,s_{3})a^{%
\ast}(p,s_{3})+e^{-ipx}v^{A,\dot{B}}(p,s_{3})\cdot b(p))\frac{d^{3}p}{2p_{0}}
\label{int}
\end{equation}
where the intertwiner functions $u^{A,\dot{B}}(p,s_{3})$ and their
charge-conjugate counterpart are $(2A+1)(2B+1)$ component which intertwine
between the unitary $(2s+1)$-component Wigner representation and the
covariant $(2A+1)(2B+1)~$dimensional spinorial representation labeled by the
semi-integer $A,\dot{B}$ which characterize the finite dimensional
representations of the covering of the Lorentz group $SL(2,C)$. The
intertwiner functions are determined in terms of group theoretic properties;
the use of modular localization is not necessary.

There is one annoying loophole in this construction in that the important
massless vector potential and more generally tensor potentials do not exist
in a point-local form since they violate positivity \footnote{%
A similar problem exists for massless fermionic fields for $s\geq3/2.$}. In
that case the way out has been to quantize classical gauge theory. The
problem with this is that Hilbert space positivity, which is classically
irrelevant but indispensible for the probabilistic properties of quantum
theory, is violated in the quantized result. It can be recovered only for
the gauge invariant part of the theory which excludes the important matter
fields but includes the local observables in the form of gauge invariant
fields. There exist no perturbative approach based on point-local fields
which is able to avoid the use of unphysical fields. This requires the
introduction of additional indefinite metric degrees of freedom and ghost
fields which have no counterpart in the classical theory but are necessary
to implement the operator gauge transformations; the latter bear no relation
with physical symmetries but are nevertheless needed in order to extract the
physical quantities from an unphysical setting.

The physical reason for being forced to take recourse to quantum gauge
theory is that there is a clash between positivity and localization of which
the problem with massless pl free vector potentials is only the tip of an
iceberg. It also manifests itself in the nonexistence of massless conserved
pl currents for $s\geq1$ as well as that of massless energy-momentum tensors
for $s\geq2$ \cite{WW}. In the presence of interactions its manifestation
affects even massive QFTs in that it is the cause of the nonexistence of
positivity preserving renormalizable interactions involving $s\geq1$ fields.
Instead of combatting this phenomenon by short distance improvement
resulting from compensations of part of the positive probability with
negative metric contributions the positivity preserving way of improving
short distance scale dimensions of fields is a relaxation on tightness of
causal localization by passing from pl to sl free fields.

\section{String-localized fields}

Point-local free fields for spin $s$ or helicity $h~$are uniquely determined
in terms of their covariance. Massive tensor fields of spin $s$ have short
distance dimension $d_{sd}=s+1$. Interaction densities $L$ are defined in
terms of Lorentz-invariant Wick-ordered products of free fields and
according to the power counting bound of renormalizability $d_{sd}(L)\leq4~$%
there are no renormalizable interactions with $s\geq1.$ Positivity obeying
massless point-local tensor fields of helicity $h\geq1$ and tensor degree $h 
$ do not exist; this is a consequence of the absence of intertwiners from
massless helicity $h$ unitary Wigner representations to ($h/2,h/2$)
covariant tensor fields.

Both problems are related to the positivity of pl fields which is in turn a
consequence of the unitarity of Wigner's representation theory. For $s=1$
this problem is formally solved by replacing the Hilbert space by a
positivity-violating indefinite metric Krein space which lowers the $d_{sd}$
of the Proca field from $2$ to $1$. The indispensable positivity property is
then recovered for the subtheory of local observables (which includes field
strengths) whereas the important charge-carrying fields, which relate the
causality principle with particles, remain outside gauge theory. Hence gauge
theory, although in its classical form a complete theory with local gauge
invariance, is an incomplete QFT in which gauge symmetry plays the role of a
formal device whose only purpose is to filter the physical subtheory from an
unphysical (negative probabilities containing) description.

The new string%
\'{}%
local field theory (SLFT) is a complete QFT whose construction is based on
the observation that the culprit for the indefinite metric and the resulting
lack of positivity of probabilities is the use of covariant pl fields. As
soon as one uses their covariant sl siblings in a way which is consistent
with their weaker localization one is led to the beginnings of a new
renormalization theory in which $s=1$ (and more generally $s>1$) fields have
a spin-\'{\i}ndependent short distance dimension ($d_{sd}(bosons)=1,$ $%
d_{sd}(fermion)=3/2$) and thus permit the formation of tri- or quadri-linear
interaction densities $L$ with the power counting bound (pcb) $%
d_{sd}(L)\leq4.$

The "naturalness" of sl fields follows from a theorem in LQP \cite{Haag}
which states that in the presence of a mass gap there exists for each
particle type an interpolating sl field\footnote{%
In the algebraic setting of LQP this corresponds to an interpolating
operator which belongs to an algebra of an arbitrary narrow space.like cone
(whose core is a space-like string).}. From SLFT one knows that interactions
containing $s\geq1$ fields lead (apart from pl observables) to sl
interacting fields. The terminology QFT and in particular LQP always refers
to positivity maintaining descriptions; indefinite metric decriptions will
be referred to as gauge theory (GT).

The Wightman setting of QFT and Haag's LQP cannot dispense with positivity;
its absence does not only affect the probability interpretation but gauge
dependent fields also fail to describe the correct causal localization. In
order to solve the positivity problem it is important to understand the
relation between tightness of localization and short distance dimensions in
more detail. Starting from a massive $d_{sd}=2\ $Proca vector potential $%
A^{P}$ it is easy to see that the covariant string-local solution of the
operator-valued differential 2-form $F_{\mu\nu}=\partial_{\mu}A_{v}^{P}-%
\partial_{\nu}A_{\mu}^{P}$%
\begin{equation}
A_{\mu}(x,e)=\int_{0}^{\infty}F_{\mu\nu}(x+\lambda e)e^{\nu}d\lambda ,\ \
e^{2}=-1   \label{s}
\end{equation}
Here the linear form of the space-like string and the Lorentz transformation
of $e\ $in the covariant secures the covariance of sl fields\ and the
integration to infinity insures the lowering of the dimension from $%
d_{sd}=2\ $to $1.\ $

By starting from a massive general spin $s$ tensor field and forming the
field strength which corresponds to a $2s$-form the $s$-fold repetition of
the line integration results in a $d_{sd}=1$ $e$-dependent string-local
counterpart while their iterative application to the pl degree $s\ $tensor
potential define the $s\ $tensorial escorts of maximal degree $s-1\ $ \cite%
{E-M}. A similar idea applied to the point-local spinor-tensor potential of
half-integer spin $s\ $(one spinor- and $s-1/2$ $\ $tensor indices) leads to
a similar situation in which the resulting string-local spin $s$ field has
the same dimension as a $s=1/2$ Dirac field namely $d_{sd}=3/2,\ $%
independent of $s.$ By taking a more general integration measure $%
d\lambda\rightarrow \kappa(\lambda)d\lambda$ one can vary the $d_{sd}$
continuously down to zero.

Only the so constructed massive sl tensor potentials have smooth massless
limits\footnote{%
What is meant is that the 2-point massive correlation functions converge to
those of the massless helicity fields but the representations of the
operators of course remain unitarily inequivalent}. This does not only lead
to the sl replacement of the missing pl massless potential but it also
defuses a No-Go theorem by Weinberg and Witten claiming that massless
energy-momentum tensors do not exist for $s\geq2\ $\cite{WW}. The correct
statement is that \textit{pl conserved massless E-M tensors do not exist};
they have to be replaced by \textit{sl E-M tensors which are different as
densities but lead to the same global charges} (generators of the Poincare
group).

One may think that the use of sl instead of pl fields which converts pcb
violating pl interaction densities into pcb obeying sl ones renders a model
renormalizable. But as often in QFT, trying to patch up a problem creates
another problem at an unexpected place.. Without the fulfillment of an
additional condition, which prevents the total delocalization at higher
orders, the validity of the pcb is insufficient to guaranty consistency.
This additional requirement will be addressed in section 6.

Covariant sl fields can be constructed in a rather elementary way from their
pl counterparts without referring to the more foundational LQP. But
overlooked simple constructions arise sometimes in a roundabout way. The
study of sl fields did not start in the above form but rather developed in
the aftermath of solving the foundational problem, more than 7 decades old,
of the causal localization of Wigner's infinite spin matter for which it was
essential to use modular localization theory.

In \cite{BGL} it was shown that all positive energy representations are
localizable in arbitrary narrow (noncompact) space-like cones. Since it is
well known that the massive and finite helicity zero mass class is pl
generated and that the generating fields of the infinite spin
representations cannot be pl Wightman fields \cite{Y} it seemed likely that
their generating fields are localized on the semi-infinite string-like cores
of a space-like cones. Using the modular localization of the LQP setting it
was possible to construct the intertwiner functions $u(p,e)$ which relate
the momentum space Wigner creation/annihilation operators with covariant sl
fields \cite{MSY}. Previous attempts in terms of Weinberg's group theoretic
method based on covariance had failed.

Meanwhile there appeared a rather sophisticated direct proof which excludes
the existence of nontrivial compact modular localized Wigner-Fock subspaces 
\cite{MLR}. It uses the spatial version of modular localization (the $K$%
-spaces) sketched in the previous section . This raises the question about
possible physical properties of those fields. The new setting of sl
perturbative renormalization theory strongly suggests that this infinite
spin matter is inert with respect to interaction with normal matter. Matter
which only exists in the form of free fields and, through the use of its
energy momentum tensor in Einstein-Hilbert equations, may lead to
backreactions on the gravitational field, is an interesting candidate for
dark matter since its "coldness" is natural \cite{dark}

The same method of modular localization applied to Wigner's unitary
representations of ordinary matter led to the rather large class of massive
and finite helicity massless sl fields which also can be directly
constructed in terms of semi-infinite line integrals over pl fields.

The next section addresses the question of interest for many readers \textit{%
are string-localized fields related to ST theory?}.

\section{21st century physics, or, the new phlogiston?}

The naturalness of string-localization in LQP and its generic appearance in
all positivity preserving renormalizable interactions involving $s\geq1$
particles begs the question of its relation to string theory (ST).

To understand how particle physicists arrived at ST it is helpful to recall
its historical roots which can be traced back to ideas about an autonomous
S-matrix theory. This refers to attempt to formulate a theory of scattering
amplitudes without the use of its large time asymptotic relation with QFT.
The problem of such a project is that causal localization principles of QFT
are not available in such a direct construction of global on-shell objects.

A possible way out was to look for analyticity properties which generalize
those of the dispersion relations. The most conspicuous model-independent
property of on-shell restrictions of Feynman diagrams is the \textit{%
analytic crossing property}. In order to separate this property from its
(for strong interactions useless) perturbative context Mandelstam proposed a
representation for the elastic scattering amplitude which incorporates such
a property.

The historical origin of ST cannot be understood without Veneziano's
subsequent Dual Model which replaces Mandelstam's representation by a
concrete crossing symmetric meromorphic functions which substitutes the
elastic scattering continuum by a trajectory of particle poles.

ST started by viewing such on-shell particle mass trajectories as
manifestations of strings in spacetime in analogy with the energy spectrum
of a chain of quantum mechanical oscillators. This new spacetime
interpretation implied a return to an off-shell description based on the
quantization of actions of world sheets traced out by strings in spaetime.
The interaction between such strings was assumed to be described in terms of
splitting and recombining tubes representing world sheets.

The positivity requirement on this quantization selected one model in which
the spacetime was the target space of a certain $d=1+1$ conformal field
theory associated to a 10-component supersymmetric current. Our
4-dimensional world was to result from a Kaluza-Klein dimensional reduction.

Since Haag's LQP\textit{\ comprises all models which fulfill the causal
localization principles in a (positivity obeying!) Hilbert space} \textit{%
setting} and ST falls according its protagonists into this category, the
obvious question is whether the objects of ST describe, as string theorists
claim, string-localized objects in the sense of causal localization in
Minkowski space. If localization in ST really means what the terminology
suggests, two string operators should commute if the strings are spacelike
separated (the quantum version of Einstein-causality); there is no other
physical meaning which one can attribute to quantum strings localized in
spacetime.

Freed from a quantization parallelism to classical physics, the LQP
formulation is synonymous with a realization of causal localization
principles in the context of quantum theory which means in particular that 
\textit{string-local operators are defined as objects in spacetime which are
causally localized i.e. two string operators commute if they are relatively
spacelike separated}. \ 

Causal localization is inexorably connected to vacuum polarization and the
strength of the vacuum polarization clouds depend on the tightness of
localization. This affect in particular the short distance scale dimensions
of pl fields. If the alleged "stringy" objects of ST bear any relation to
spacetime strings they must be related to the sl fields of LQP even if they
had been constructed in a different way from that of sl fields. The main
point of contention is whether the objects of ST are really string-local in
any with relativistic causality compatible sense.

In order to understand that string theorists use the terminology "string"
for something which bears no relation with localized quantum objects in
spacetime it is helpful to look at what they are doing and understand why 
\textit{they} think they are addressing propertie of quantum localization.
Before addressing the quantization of the Nambu-Goto action or constructing
their 10 dimensional "superstring model" from the action of a particular 10
component supersymmetric $d=1+1~$conformal current model string theorists it
is helpful to take a critical look at their view of the quantum theoretical
counterpart of particle world lines \cite{Pol}.

The model is defined in terms of the relativistic action $\sqrt{-ds^{2}}$but
the resulting covariant classical world line has no quantized counterpart
since particle operators $\vec{q}(t)$ only exist in (nonrelativistic)
quantum mechanics (the nonintrinsic "Born localization") and the quantum
theoretical description of a single relativistic particle uses Wigner
representation theory. From the latter one can construct free fields which
and the point-local free fields can be reformulated in terms of a
relativistic action. There is simply no access to wave functions of
relativistic particles in terms a quantization of actions describing
relativistic world lines and hence this construction turns out to be a squib
load.

The theory which describes relativistic particles is Wigner's construction
of \textit{unitary representations of the Poincar\'{e} group which cannot be
accessed by quantization of classical actions}; in fact his 1939 unitary
representation theory was the first successful \textit{intrinsic} quantum
construction of a relativistic particle theory. As we know nowadays this
theory already containes the germ of causal localization\footnote{%
Wigner tried ito find a representation theoretical signal of causal
localization and became disappointed when he realized that the
"Newton-Wigner localization" did not solve the problem \cite{rem}. The
conceptual prerequisites for the later "modular localization did not exist
at that time.} in the form of modular localization of positive energy states
which is closely related to the causal localization of fields.

Only on this level of causal localization of fields can one make contact
with the quantization of pl fields (section 3). The more generic and
important covariant sl fields cannot be accessed in this way (section4).
They are objects which are pure quantum in that the umbilical cord of an
alleged quantization parallelism has been cut. This is our main motivation
for giving much space to causal localization in an article dedicated to the
memory of the protagonist of LQP which places causally locaiized operator
algebras into the center stage.

This leaves the question of what remains of ST if it is not a theory of
quantum strings in spacetime. An authoritative answer from somebody who has
spend a good part of his professional life to understand the physical
content of the Nambu-Goto action is that it describes an infinite set of
conserved charges as one finds in $d=1+1\ $integrable QFTs. But different
from integrable d=1+1 QFT there is no trace of anyspacetime localization in
N-G models \cite{Pohl}.

The fusion and splitting of world sheets as a description of spacetime
strings in analogy to the interpretation of perturbative Feynman graphs as
coalescing and splitting of point-like particles represents an attempt of
string theorists to create localized interactions in terms of classical
metaphors. On the other hand the fact that this is based on
misunderstandings of quantum causal localization does not invalidate the
mathematical use of such constructions as an inspiration for interesting
topological, algebraic and geometric constructions. ST also led to some new
computational techniques which are useful in other areas of particle theory.
If it did not prevent careers by occupying many research position and
distract many from problems of particle theory it would be easier for
physicists to appreciate its mathematical contributions.

One reliable result which was obtained by string theorists, although not
related to string localization, is a theorem by Brower \cite{Brow}. It
states that the irreducible \textit{superstring algebra,} defined in terms
of the aforementioned supersymmetric 10 component conformal field theory,
carries a positive energy Wigner representation which decomposes into a an
infinite direct of sum of irreducible ($m>0,s$) and ($m=0,h$) Wigner
representations.

This has an interesting connection with an old project by Majorana. In
analogy to the description of the discrete spectrum of the hydrogen atom in
terms of a $O(4,2)$ representation, Majorana's idea was to construct a group
algebras of a higher dimensional group which contain a tower of particle
wave function spaces. This idea underwent a revival in the 60s in the form
of "dynamical groups" leading to a discrete spectrum of particles. Apart
from the fact that the irreducible superstring algebra associated to the
conformal field theory is not a group algebra, Brower's theorem is similar
in that it refers to a particular particle spectrum which originates from
the action of the Poincar\'{e} group on an irreducible algebra.

In the eyes of string theorists the map of the two-dimensional conformal
space into the 10 dimensional target space describes what they call a string
in form of a world sheet defined in terms of a map from the two-dimensional
conformal space into the 10-dimensional "target" spacetime. Without these
"string glasses" one only sees a discrete direct sum of unitary Wigner
representation (but no target space localization) whose conversion into
covariant free fields leaves the choice of pl or sl. As for any unitary
positive energy Wigner representation which carries a modular localization
structure it is the interaction which decides about the localization:
renormalizable interactions of $s<1$ require the use of pl fields whereas
renormalizability and positivity in the presence of $s\geq1~$fields can only
be maintained in terms of string-localization.

The problem of localization of fields has nothing to do with that particles;
the latter remain what they always were: states described by in time
dissipating Wigner wave function which, as a result of their positive energy
content, do not admit a causal pl or sl localization. What may be idealized
as a pl event is the registering in a counter; the difference whether the
fields whose application to the vacuum create these states were sl or pl
only manifests itself in a more spacetime spread out region of clicks. It is
important to have a clear view of the relation between fields and particles
in order to understand Haag's stomach ache with string theorists view of
strings and particles (see below).

The heuristic picture of ST in terms of splitting and recombining world
sheets has led, particularly in the hands of Ed Witten, to highly
interesting new ideas and results in geometry and topology but this has not
helped to its \textit{physical} use. Concepts as that of \textit{modular
localization} which are the raison d'\^{e}tre for local quantum physics have
remained outside ST and its derivations (extra-dimensions, AdS-CFT).

The promise to address the issue of string-localization (which is the origin
of the terminology ST) has remained unfulfilled and there is no way in which
this can change. It is simply not possible to create a new theory without a
foundational dispute with the in every aspect successful and comprehensive
QFT. Mathematical enrichments cannot hide the fact that the physical
contributions of ST to particle theory has remained smaller than any
preassigned epsilon.

Historian of physics who would seriously attempt to take stock of viable
theoretical physics concepts which originated within 50 years of ST will
presumably have a hard time to account for what had been achieved. Haag's
reaction to the present situation in this respect is quite interesting.

On the occasion of presenting a seminar talk more than three decades after
having held a visiting position at the university of Princeton, he was hard
pressed by Ed Witten to join the ST community. Haag's recollection of this
situation can be found in his published reminiscenses \cite{rem}. He writes:

"I visited Princeton in the early 90ies. At that time Sam Treiman was head
of the physics department at the university. I had known him since 1958 and
highly appreciated his sober judgement. So I asked him about his assessment
of the future of string theory. He said that he had not occupied himself
with it but that he was supporting it without reservation because the people
who worked on it were very very good. He meant primarily Ed Witten who was
now the spearhead of this approach. I had been asked to give a physics
colloquium talk about my views on quantum gravity and hoped to have some
discussion with Ed Witten. Next morning he greeted me by saying:
\textquotedblleft Your talk was very interesting but I would really advise
you to work on string theory\textquotedblright. When he saw the somewhat
incredulous look on my face he added \textquotedblleft I really mean it. I
shall send you the manuscript of the first chapters of our
book\textquotedblright. This ended our discussion. Back in Hamburg I
received the manuscript but it did not convert me to string theory. I
remained a heathen to this day and regret that meanwhile most physics
departments believe that they must have a string theory group and have
filled their vacant positions with string theorists. To be precise: It is
good that people with vision like Ed Witten spend time trying to develop a
revolutionary theory. But it is not healthy if a whole generation of young
theorists is engaged in speculative work with only superficial grounding in
traditional knowledge."

Haag's critical comments should be seen in the context of his conduct of
research which is distinguished by self-reliance and a self-critical
scrutinizing. This may have had its origin in the circumstances in which his
interest in theoretical physics arose. As a teen ager he was on a private
visit to his sister who lived in the UK when in 1939 the war started and he
could not return to his mother in Stuttgart and finish high school (in 1939
his age was 17). As a German citizen he was shipped to Canada where he spend
the years of war in a detention center. There he managed to get hold of a
book on physics which he used for self-studies.

Returning at the end of the war from the Canadian camp to Stuttgart he found
himself in a war-devastated city without a functioning academic teaching
program. In such a situation self-reliance and intellectual autonomy were
important.

To find his former strongly independent minded colleague from Princeton Sam
Treiman three decades later in a state of dependence on authorities
concerning a subject which he considered of prime importance was apparently
somewhat unexpected for Rudolf Haag.

The reaction of Haag to both Witten and Treiman can be best commented in
form of a metaphor: it is not enough to believe to have discovered the Lapis
Philosophorum, one must also have the charisma to convince sufficiently many
prestigious persons to share this belief.

Haag was hardly impressed by mathematical work whose motivation did not
originate from fundamental physical problems. The situation of QFT after the
discovery of perturbative QED, in which different prescriptions of
"exorcising infinities" amazingly led to mutually compatible results,
certainly motivated him to look for a more coherent description which
finally culminated in his framework of what he later referred to as local
quantum physics (LQP). He was convinced that all properties of causally
localized quantum matter was encoded in the relation between observable
algebras $\mathcal{A(O})~$labeled by their spacetime localization region $%
\mathcal{O}$. His innovative strength resulted from his ability to find the
appropriate mathematical setting for his physical ideas. For more detailed
mathematical knowledge he relied often on mathematically more knowledgeable
collaborators.

The mere fact that ST did not arrive at any observationally testable
proposal throughout its 50 years of existence (which is its common critique)
was of not much concern to Haag. One can assume that both he and Witten
shared the belief that foundational ideas should have all the time they need
to evolve.

He would however have expected that the exploration of a foundational idea
should lead to a steady increase of knowledge about important theoretical
problems of particle physics. His LQP led to a profound understanding of why
the local structure of QFT is much more powerful than its classical
counterpart. Together with his collaborators Sergio Doplicher and John
Roberts he derived a classification of internal symmetries and the absence
of parastatistics as consequences of properties of superselection rules
which in turn were obtained from localization structure of local
observables. Theorems in Wightman's formulation of QFT \cite{St-Wi} have
their counterpart in LQP and some properties (including the causal
completion property and Haag duality) permit no natural formulation in
Wightman's field theoretic setting. Most of the results were obtained by a
few researchers; the number of people working on foundational problems of
QFT in the 80s was rather small compared with that in ST.

On the other hand ST led to "hot topics" which produced thousands of
publications and provided university positions to their authors who in many
cases obtained their positions because they were working on such a
fashionable topic.

The formation of such transient fashions is reminiscent of bubbles in the
financial market but it is presently not clear to me whether this is the
consequence of the increasing dominance of financial capitalism in all areas
of life or whether this has its more specific explanation in the seducing
charisma of the protagonists of ST; probably it is the result of a Zrizgeist
in which both interplay.

An example of such a bubble in the wake of ST is the physical use of the
mathematically correct AdS-CFT isomorphism. Fronsdal observed already in the
60s that the spacetime symmetry group of the so-called anti-de Sitter
spacetime is isomorphic to the symmetry group of a one dimensional lower
conformally invariant spacetime. As a result of a presumed connection
between five-dimensional gravity with gauge theories in four spacetime
dimensions the problem of a possible QFT isomorphism behind the AdS-CFT
group theoretical relation returned in the late 90s.

Since the mismatch between degrees of freedoms in comparing QFTs in
different spacetime dimensions renders the use of fields unsuitable, the
existence of a AdS-CFT isomorphism was finally rigorously established within
the algebraic LQP setting \cite{Re}. The proof showed that the mismatch of
the cardinality of degrees of freedom between isomorphically related QFTs in
different spacetime dimension only affects the causal completeness property,
which is the quantum counterpart of the hyperbolic propagation of classical
Cauchy data.

As mentioned before (section 2) this problem is not limited to this
particular isomorphism but it affects all problems related to
"transplanting" quantum matter between spacetimes of different dimensions; \
metaphorically speaking the resettling from higher to lower dimensions
suffers from overpopulation whereas in the opposite direction it causes
"anemia" in that there are not enough degrees of freedom to sustain AdS
fields. The most appropriate way is to express the isomorphism in terms of
localized operator algebras \cite{Re}.

Methods based on quantization of actions are not suitable for the study of
such isomorphisms because the notion of cardinality of degrees of freedom
has no counterpart in classical or semiclassical field theory and therefore
tends to be overlooked. This mismatch of cardinality of degrees of freedom
removes the rug from underneath the idea of \textit{extra dimensions}.

This situation reveals a dilemma of present foundational theoretical
research. The increasing number of researchers in particle theory does not
seem to lead to a broadening of topics and an increase of critical
knowledge: it rather tends to favor monocultures and a loss of past
knowledge and wisdom.

It is interesting to compare the present situation with that which Haag met
in the 50s during his stay at the Niels Bohr Institute in Copenhagen (\cite%
{rem} page 269) and which still dominated the scientific discourse during
the 60s. This was the high time of the "European Streitkultur" in which
different views about problems and the elimination of incorrect or
misunderstood ideas as well as what new directions to take was hammered out
in often heated and sometimes even polemic disputes between equals.

After the discovery of quantum theory in Europe different univerities often
represented different schools of thought ("the Copenhagen interpretation")
which led to rivalries and sometimes even to polemics between the
protagonists of these schools. When the political situation worsened many of
the leading scientist left for the US and this rivalry spread to the US. In
the 50's and 60's the discourse was dominated by individuals as Pauli, Jost,
Kallen, Landau, Lehmann, Feynman, Schwinger to name just a few.

A positive effect of this often somewhat rough way of communicating was that
futile or erroneous ideas (the S-matrix bootstrap, peratization,
Heisenberg's spinor theory, Reggeology, SO(6),..) could not survive for more
than a decade. In highly speculative research as particle theory the
occurrence of wrong turns is inevitable and therefore the existence of a
lively "Streitkultur" is important. In such a climate the survival of theory
bases on misunderstood or evenfor more with a Nearly all our important
theoretical results and computational tools, which later became household
goods, originated in those times.

Compare this with the legacy of 5 decades of ST; apart from some new
calculational techniques and an enrichment of certain ares of mathematics it
is hard to find any remaining contribution to particle theory. ST and its
legacy appears increasingly as a gigantic bubble in particle theory which
has led to extra dimensions, branes, M-theory,...which contradict basic
properties of QFT. The more damaging legacy of this bubble is the incorrect
view of the field-particle relation which ignores previously gained wisdom.

It is interesting to quote Haag on this matter \cite{rem}. "In many
popularized presentations the starting point of string theory is explained
as the replacement of the fundamental notion of "particles" with its
classical picture of a point in space or a world line in spacetime by a
string in space respectively a sheet in spacetime. This, I think, is a
misunderstanding of existing wisdom. First of all, paraphrasing Heisenberg ,
one may say "Particles are the roof of the theory, not its foundation."
Secondly points in space cannot be defined as the position of particles in a
relativistic theory."

The understanding of the relation between fields and particles is one of the
most important and subtle achievements of the 50's and 60's. Quantum fields
are the carriers of the foundational causal localization principles and are
generally not objects of direct observations\footnote{%
However their quasiclassical approximations (usually expectation values in
coherent states) are measurable in QED (the massless photons are important).}%
.

In particular the correct formulation of string-like localization of fields
does not imply that particles become "stringy". Covariant free fields exist
in pl as well in a sl form; applied to the vacuum pl and sl fields create
states in the same Wigner representation. Their difference only shows up in
interactions; in particular renormalizability requires to use sl $s\geq1$
fields in the interaction density (section 6) and higher order interactions
transfers this sl localization to the originally (in lowest order) pl $s<1$
fields.

There \textit{naturalness} of sl localization is supported by a theorem (%
\cite{Haag} section IV.3) which states that in models with a mass gap and
local observables the asymptotic particles and their scattering matrix can
be described in terms of the large time asymptotic behavior of operators
which are localized in arbitrary narrow spacelike cones whose cores are
spacelike strings. Taking this theorem from its algebraic LQP setting to
that of QFT formulated in terms of covariant fields it states that in such a
theory the Wigner particles can be described in terms of interpolating
covariant sl fields.

What was not known at the time when Haag wrote his reminiscences was that
positivity violating local gauge theory can be reformulated in terms of a
positivity obeying sl theory and that the idea of positivity preservation
requires the use of sl $s\geq1$ fields in all interactions involving such
fields. Viewing local gauge theory as a prick in the flesh of QFT he
certainly would have appreciated this recent insight. It strongly suggests
that sl is the standard situation and interacting pl fields are limited to $%
s<1$ interactions.

The state space generated by charge-carrying fields coupled to photons has a
more complicated particle structure ("infraparticles") than a Wigner-Fock
particle space and its description remains essentially unknown, although
there exist efficient momentum space descriptions for photon-inclusive cross
sections (Bloch-Nordsiek, Yenny-Frautschi-Suura \cite{YFS}). The loss of
foundational knowledge on the road to a "theory of everything" bodes ill for
the future of particle theory.

Research at the frontiers of particle theory is an intrinsically highly
speculative intellectual activity. According to one of Feynman's allegorical
comments it is sometimes necessary "to dive into the blue yonder" but, as he
continues to point out, such jumps should be only undertaken from a platform
of solid knowledge of QFT, so that one can return and try other directions
instead of getting lost for the rest of one's life in a hopeless project. In
his last years of his life he saw the problems originating from the
popularization of ST but he was unable to influence its course.

For the first three decades of post-renormalization QFT it was possible to
make important discoveries without deep conceptual investments. \ With some
basic knowledge about computational techniques of QFT and a heuristic
understanding of the field-particle relation one could make important
discoveries "by pulling up one's sleeves" and starting a calculation and, if
necessary, correcting it or trying other directions. \ 

It was not important whether a consistent and interesting-looking result was
derived from a fully correct theory since there was always the possibility
to consider incomplete or faulty theoretical ideas which led to important
discoveries as a temporary placeholder and hope for a future more
appropriate understanding. In this way Dirac discovered antiparticles within
the less than correct hole theory.

This way of conducting research was exhausted at the end of the 70s. ST and
its derivates are the result of attempts which tried to extend this success
without making new conceptual investments. According to Phil Anderson the
overwhelming success of particle theory in its first decades of existence
had created a kind of intellectual arrogance about Nature. It was easier to
speculate about how to go beyond QFT and claim to arrive at a theory of
everything than to do the hard work necessary for the understanding of the
deeper conceptual layers of our most successful and comprehensive theory of
the material nature of the world. The superficial image of QFT which the
leading influential representatives of ST painted and transmitted was that
of "old QFT" being replaced by ST.

The title of this section contains the term "phlogiston" which in pre-oxygen
times represented a substance which allegedly escapes in the process of
burning. The phlogiston theory only disappeared when Lavoisier at the time
of the French revolution discovered oxygen and its role in combustion. ST
cannot disappear in this way because unlike phlogiston it has no observables
consequences. As long as there are renown scientists (including bearers of
Nobel prizes) among its protagonists it will persist. The times in which it
was possible to clarify issues in disputes between equals as in the old
European Streitkultur are long gone.

\section{String-local perturbation theory}

In the absence of interactions massive pl fields and their sl counterparts
are two physically equivalent ways of coordinatizing a free LQP model; in
particular the sl fields maintain the asymptotic relation between fields and
particles which is the basis of time dependent (LSZ, Haag-Ruelle) scattering
theory. The Wigner-Fock particle structure breaks down in case the
interaction involves massless $s\geq1$ fields.

Haag's LQP and in particular the concept of modular localization played an
important role in raising awareness about the important role of covariant sl
fields and led to their first constructions \cite{MSY}. It turned out that,
apart from fields associated to Wigner's infinite spin representation class,
all covariant sl free fields can be directly constructed in terms of pl
free.weighted line integrals over pl fields.

Perturbative constructions involving sl fields are very much in their
infancy, and if the following simple illustrations encourage other particle
theoreticians to engage in the exploration of this extremely rich area of
research this section will have accomplished its purpose. The only
perturbative interactions which have been considered up to date are
couplings of massive $s=1$ vector potentials to lower spin ($s=1,1/2$)
matter fields.

In order to pass from a nonrenormalizable pl interaction density to its less
singular sl counterpart one needs a linear relation between the $s\geq1$ pl
fields and their less singular sl counterparts. For massive vector
potentials this relation reads 
\begin{align}
A_{\mu}(x,e) & =A_{\mu}^{P}(x)+\partial_{\mu}\phi(x,e),\ \ \phi(x,e)=\int
d\lambda e^{\nu}A_{\nu}^{P}(x+\lambda e)  \label{rel} \\
A_{\mu}(x,e) & =\int d\lambda e^{\nu}F_{\mu\nu}(x+\lambda e),\ \ F_{\mu\nu
}(x)=\partial_{\mu}A_{\nu}^{P}-\partial_{\nu}A_{\mu}^{P}  \nonumber
\end{align}
It involves an additional \textit{scalar} sl $\phi$ field.

Scalar covariant sl fields for any integer spin have been first constructed
in \cite{MSY}: their semi-integer fermionic counterparts are sl Dirac fields
for any halfinteger spin. They violate the connection between physical spin
and the covariant transformation property of their pl counterparts. It turns
out that only those sl fields which appear in a linear relation between
covariant pl fields and their sl counterparts as in (\ref{rel}) play an
important role in the new SLFT renormalization theory. On a purely formal
level their appearance in the form of a gauge transformation is reminiscent
of scalar negative metric Stueckelberg fields which appear in the operator
gauge transformation between the Feynman gauge and its unitary counterpart;
they convert the renormalizable but unphysical matter fields into their
formally physical but very singular\footnote{%
They are not polynomially bounded and hence they cannot be Wightman fields.}
counterparts \cite{L-S} \cite{Scharf} \cite{Ruegg}.

The conceptual and mathematical situation in (\ref{rel}) is very different.
The three linearly related fields live on the same Wigner Fock space of $s=1$
particles and belong to the same sl localization class (i.e. they are
relatively Einstein-causal in the sense of string-localization). The sl
setting avoids the introduction of additional (unphysical) degrees of
freedom and in this respect may be viewed as the result of the "application
of Ockham's razor to gauge theory" \cite{E-M}. It is not a gauge theory
because the local operator gauge transformations (different from the global
U(1) transformations) cannot be defined without the presence of additional
(indefinite metric) degrees of freedom.

The computational tests and the conceptual coherence of the new SLFT setting
leave no doubt that after more than 70 years one finally arrived at a new
setting which reunites the $s<1$ renormalizable interactions with those of $%
s\geq1~$under the same conceptual roof of causally localized and positivity
obeying (quantum probability preserving) genuine quantum theories.

The sl scalar $\phi$-fields will be referred to as an "escort" of the pl
Proca potential. Only the correlation functions of the sl potential permits
a massless limit whose reconstructed Wightman field \cite{St-Wi} is the
vector potential of Wigner's $h=1$ helicity representation\footnote{%
Since massless Wigner representations are unitarily inequivalent to massive
ones, the mooth behavior refers to the expectation values and not to the
operators.}. The purpose of this care about massless limits, which in the
present context appears pedantic, is to raise awareness about the
reconstruction problem of a physical Hilbert space which corresponds to the
Wigner Fock space provided by scattering theory in the presence of a mass
gap. One knows very little about a particle like description of this limit
(the problem of \ "infraparticles" and confinement).

The linear relation (\ref{rel}) between $A^{P},A$ and the escort $\phi$ is
really a linear relation between their intertwiners. Computing the
intertwiners $u_{\mu,s_{3}}(p,e)$ of $A_{\mu}(x,e)$ and $u_{s_{3}}(p,e)$ of$%
\ \phi(x,e)\ $in terms of of the intertwiner $u_{\mu,s_{3}}(p)\ $of$~A^{P}~$%
\[
A_{\mu}^{P}(x)=\frac{1}{(2\pi)^{3/2}}\int\sum_{s_{3}}e^{ipx}u_{%
\mu,s_{3}}(p)a^{\ast}(p,s_{3})+h.c. 
\]
using their definition in (\ref{rel}), one verifies the linear relation%
\footnote{%
The Fourier transforms of the Heaviside funtions $\theta(x)$ account for
denoinators $1/pe.$}; for general spin the corresponding formula contains $s$
escorts \cite{E-M} \cite{beyond}. A more geometric interpretation views the
escort field in the context of the Poincare lemma applied to the
differential 2-form $F_{\mu\nu}$.

It is interesting to note that the massless limit preserves the number of
degrees of freedom: 2 are accounted for by $h=1$ and one is carried by the
massless limit of the pl scalar field $\phi^{P}(x)=\lim_{m\rightarrow0}m%
\phi(x,e).$This prevails for spin $s$ tensor fields for which the linear
relation (\ref{rel}) contains $s$ tensorial escort fields.

Starting from the 2-point function of the unique positivity-obeying
long-range massless vector potential and "switching on" the mass one cannot
return to the short range 2-point function of the Proca potential without
the escort $\phi\ $plying its (\ref{rel}). The difference from the Higgs's
mechanism is that escort fields do not introduce new degrees of freedom; so
whenever the presence of an additional Higgs field is necessary it must be
for other reasons.

The important property of the sl vector potential $A_{\mu}(x,e)$ can be seen
in its$\ $2-point function\footnote{%
All pl fields have polynomial two-point functions in p.}%
\begin{align}
& \left\langle A_{\mu}(x,e)~A_{\nu}(x^{\prime},e^{\prime})\right\rangle =%
\frac{1}{(2\pi)^{3/2}}\int e^{-i(x-x^{\prime})p}M_{\mu\nu}(p;e,e^{\prime })%
\frac{d^{3}p}{2p_{0}}  \label{2-point} \\
& M_{\mu\nu}=-g_{\mu\nu}+\frac{p_{\mu}p_{\nu}}{pe_{-}pe_{+}^{\prime}}-\frac{%
p_{\mu}e_{\nu}}{pe_{-}}-\frac{p_{\nu}e_{\mu}^{\prime}}{pe_{+}^{\prime}} 
\nonumber
\end{align}
where the $\pm~$signs refer to the distributional boundary values $%
\lim_{\varepsilon\rightarrow0}1/(pe\pm i\varepsilon)$ from the Fourier
transforms of the Heaviside function of the semi-infinite linear string.

The gauge theoretic Feynman 2-point function (without the additional
rational p-dependent contributions) looks much simpler but contains
longitudinal positivity-violating unphysical degrees of freedom which in the
presence of interactions "infect" the matter degrees of freedom and account
for the physical limitations of local gauge theory which,~as a result of the
positivity-localization interrelation also affects causal localization%
\footnote{%
The interacting positivity-obeying electric charge carrying covariant field
of is necessary sl, even so the particle counter events can be well
localized.}.

In contrast to the Proca 2-point functions%
\begin{equation}
M_{\mu\nu}^{P}(p)=-g_{\mu\nu}+\frac{p_{\mu}p_{\nu}}{m^{2}}   \label{Proca}
\end{equation}
which has a quadratic mass divergence the sl 2-point function (\ref{2-point}%
) admits a well-defined massless limit in that it passes smoothly to its sl
helicity $h=1$ counterpart (the mass only enters the $p_{0}$).

The price for having used Ockham's razor is the appearance of nondiagonal
2-point functions of free fields, e.g. mixed 2-point function $\left\langle
A\phi\right\rangle .$ This is not surprising since both $A$ and $\phi$ are
linear combinations of the same Wigner creation/annihilation operators. This
leads to a slightly more involved perturbation theory. But it is very
worthwhile to pay this increase computational expenditure since the new
formalism does not only maintain the quantum probability but secures also
the physical localization.

Already for the free sl potentials this implies a slightly more refined
interpretation of the Aharonov-Bohm effect. It can be shown that Wilson
loops keep a topological memory of the string dependence \cite{beyond} which
leads to a violation of the Haag duality. The latter is slightly stronger
than Einstein causality ($=$ instead of $\subset$) which is intuitively
often identified with the latter\footnote{%
This is the origin of the quirky feeling about causality which makes the A-B
effect a subject of public interest.}. The violation of Haag duality is a
feature of all massless physical $s\geq1$ fields which only exist in the
form of positivity preserving sl fields.

The conversion of $d_{sd}=s+1$ potentials into their $d_{sd}=1$ sl
counterparts is a general phenomenon \cite{E-S} \cite{beyond} and has an
extension to Fermions. For instance for the massive $s=3/2$ Rarita-Schwinger
potential the corresponding escort shares the $d_{sd}=1$ with the above
scalar escort but reveals its Fermi statistics through the presence of gamma
matrices in the propagator. The claim that there exist gamma independent $%
d_{sd}=1,\ s=1/2\ $"Elko fields" is based on a misunderstanding of the
relation of free fields with Wigner's representation theory \cite{Elko}.

New physical properties arising from the reorganization of already existing
degrees of freedom into new fields represent a quite common phenomenon in
quantum mechanical many body problems. For example the Cooper pairs in
superconductivity are the result of such regrouping of electrons into
bosonic bound pairs at low temperature\footnote{%
Haag's presentation of Cooper pairs is particularly close to the spirit of
LQP \cite{Haag}.} \ Among other things they account for the change of
long-range classical Maxwell vector potentials into their short-range
counterparts inside a superconductor (F. London's screening).

In fact this analogy between escorts $\phi$ and Cooper pairs goes much
further. It clears the head from the tale about "fattening" of photons by
"swallowing" massless Goldstones and facilitates the correct understanding
why massive neutral Hermitian $H$ fields are really needed to save the
second order renormalizability of \textit{self-interacting massive vector
mesons }through short distance compensations. This is similar to what what
was expected to be a fringe benefit of supersymmetry but in the present case
it is the raison d'\^{e}tre for the $H$ field (more details below).

By analogy the long range vector potentials of photons cannot be converted
into their massive Proca counterparts by just "switching on" a mass; one
also needs the intervention of the $\phi$ escort field. QFT. Such escort
fields do not appear in renormalizable lower spin $s<1$ interactions or in
the $s=1~$indefinite metric local gauge theory\footnote{%
This is the reason why they played no role in the more than 80 years history
of QFT.}, but their presence turns out to be an indispensable aspect of
renormalizable positivity preserving LQP interactions involving $s\geq1$
particles. The conversion of $d_{sd}=s+1$ spin $s$ pl fields into their
better behaved $d_{sd}=1$ sl counterparts requires the introduction of $s$
(for half-integer spin $s-1/2$) escort fields .

Before addressing the dynamic use of sl vector potentials it is interesting
to note a relation of the massless sl potential with the $d_{sd}=1$
radiation potential. The angular integration of the sl potential over the
directions emanating from the point $x~$in a equal time hypersurface leads
to the radiation potential. Both the non-covariant radiation potential and
the covariant sl potential live in the same Hilbert space and remain
infrared convergent in the massless limit.

It has been known for a long time that the radiation (Coulomb) potential (in
contrast to vector potentials in gauge theory) lives in the Hilbert space of
the $h=1$ Wigner representation. This is the reason why investigations of
long distance (infrared) properties of charged particles have been
preferably discussed in the "Coulomb gauge" \cite{Stro}. But the radiation
potential is not a gauge in the sense in which we have used the word gauge
theory and gauge transformation in the present work since any gauge theory
needs additional (generally indefinite metric) degrees of freedom to
implement operator gauge transformation for passing from one gauge to
another.

The main reason for using the gauge theoretic setting in QED is that the
lack of covariance makes the Coulomb potential unsuitable for
renormalization. The new renormalization theory based on covariant sl fields
permits to compute the renormalized Coulomb equivalent of any operator by
angular averaging over \textit{all} string directions. This shows in
particular the equality of $e$-independent operators (local observables,
S-matrix) in both descriptions.

The guiding idea of the new sl renormalization theory is the conversion of
the power-counting bound (pcb) violating first order pl interaction density $%
L^{P}$ with $d_{sd}^{int}>4\ $into a $d_{sd}^{int}\leq4\ $renormalizable sl
density $L.$ In this way one maintains the heuristic physical content while
improving the short distance properties. This passing from pl to sl does not
affect the Hilbert space positivity (unlike gauge theory which achieves this
by a brute force compensations of part of the positive with negative metric
contributions in a Krein space setting.

Let us take a brief look at how this is done. Using the linear relation (\ref%
{rel}) one finds%
\begin{align}
& L^{P}(x)=A_{\mu}^{P}j_{\mu}=L-\partial^{\mu}V_{\mu} \\
& L:=A^{\mu}(x,e)j_{\mu},\ V_{\mu}:=\phi(x,e)j_{\mu}  \nonumber \\
& S^{(1)}=\int L^{P}=\int L   \label{ad}
\end{align}
In this way the $d_{sd}^{int}=5$ ($3$ from the $j_{\mu},\ 2\ $from $A^{P}$)
of $L^{P}$ is lowered to$~d_{sd}^{int}=4$ of $L\ \ $at the expense of $%
d_{sd}^{int}(\partial V)=5$ of the divergence term. But in models with a
mass gap this term does not contribute to the first order S-matrix in the
adiabatic limit (\ref{ad})

The problem is whether the use of the renormalizable $L$ in the adiabatic
limit extends to the higher order S-matrix and whether this idea of
adiabatic equivalence also works for the construction of correlation
functions of sl fields. Formally speaking this corresponds to the
independence from internal $e^{\prime}s$ in suitable sums of Feynman graphs
after having integrated over inner $x^{\prime}s$. For the S-matrix (i.e. in
the absence of external lines) this is reminiscent of gauge theory except
that covariant gauge fixing parameters are spacetime independent unphysical
object which bear no relation with the independently fluctuating $e$%
-directions in inner propagators.

The LQP localization theorem \cite{B-F} does not specify for which models
one \textit{needs} sl instead of pl fields; here one has to appeal to the
pcb of perturbative renormalizability criterion which reveals that there are
no positivity obeying interaction densities within the pcb limitation $%
d_{sd}^{pcb}=4$ which involve $d_{sd}=s+1$ pl $s\geq1$ fields. PCB violating
pl interactions exist only for $s<1$.

It is convenient to formulate the adiabatic equivalence in terms of the
differential form calculus of the $2+1$ de Sitter space of spacelike
directions $e^{2}=-1$ 
\begin{equation}
d_{e}(L-\partial V)=0   \label{pair}
\end{equation}
here shortly referred to as the "$L,V_{\mu}$ $pair\ condition$";\ it states
that the zero form $L-\partial V\ $is in fact exact. Its second (and
correspondingly also higher) order extension \cite{pecul} \cite{E-M}%
\begin{align}
(d_{e}+d_{e^{\prime}})(TLL^{\prime}-\partial^{\mu}TV_{\mu}L^{\prime}-%
\partial^{\mu}TLV_{\mu}^{\prime}+\partial^{\mu}\partial^{\nu}TV_{\mu}V_{\nu
})=0 &  \label{nor} \\
TL^{P}L^{P\prime}\equiv(d_{e}+d_{e^{\prime}})(TLL^{\prime}-\partial^{\mu
}TV_{\mu}L^{\prime}-\partial^{\mu}TLV_{\mu}^{\prime}+\partial^{\mu}%
\partial^{\nu}TV_{\mu}V_{\nu}) &  \nonumber
\end{align}
would be a trivial consequence of (\ref{nor}) if the time-ordered products
would not be distribution valued. It is in fact a normalization condition on
the time-ordering which extends the $e$-independence to a singular set of
intersecting strings, in particular (which include coinciding end points $x$
which carry the strongest singularities).

The higher order $L,V_{\mu}\ $pair condition can be seen as a round about
way to define time ordered products of $T$-products of pl interaction
densities whose direct calculation in the pl renormalization setting would
lead to a with the number of $L^{P}$ growing number of undetermined
parameters (second line). A higher order implementation of this formalism
requires an extension of the Epstein-Glaser renormalization theory \cite{E-G}
to sl crossings.

The pair condition (\ref{pair}) is a requirement on interaction densities.
Whereas there is no problem to satisfy the pcb condition for tri- and
quadri-linear interaction densities $L$ involving $s\geq1$ sl fields, to
construct an $L,V$ pair for a specified field content (including the
escorts) imposes restrictions on $L$ and is not always possible. But without
it the higher order perturbation would cause a total delocalization and such
a first order $L~$would not define a perturbative model of LQP.

There exists a simpler version of the pair condition which replaces the $%
V_{\mu}$ by $Q_{\mu}=d_{e}V$%
\begin{align}
& d_{e}L=\partial^{\mu}Q_{\mu},~Q_{\mu}=d_{e}V_{\mu}=j_{\mu}u,\text{\ }%
u:=d_{e}\phi  \label{Q} \\
& ~(d_{e}+d_{e^{\prime}})TLL^{\prime}=\partial^{\mu}TQ_{\mu}L^{\prime
}+\partial^{\mu}TLQ_{\mu}^{\prime}
\end{align}
Assuming asymptotic completeness (no problem in perturbation theory) the
Hilbert space in the presence of massive vector mesons is the Wigner-Fock
tensor product space of massive vector mesons with that of the $s<1\ $matter
particles. The loss of the Wigner-Fock structure in the limit of massless
vector potentials presages itself in the form of infrared divergences of the
perturbative scattering amplitudes.

The application of the pair formalism up to second order leads to
interesting new phenomena as well as new views of old phenomena. In the
following we will report on the results for four different models

\begin{itemize}
\item \textbf{Massive spinor QED}
\end{itemize}

The simplest model is massive spinor QED with $j_{\mu}=\bar{\psi}\gamma_{\mu
}\psi$; In that case the use of the standard kinematic time-ordered
propagator\footnote{%
obtained by the replacement $\frac{d^{3}p}{p_{p}}\rightarrow\frac{d^{4}p}{%
p^{2}-m^{2}}.$} yields the tree contribution to the 2-particle scattering
contribution to which we restrict our presentation. It is important to note
that the $e,e^{\prime}$ dependence is only lost in the \textit{on-shell}
S-matrix $\ $Using $e=e^{\prime}$ in off-shell relations leads to infinite
fluctuations. Since the $e^{\prime}s$ can be pictures as points on d=1+2 de
Sitter space these fluctuations are similar to those of coincident points in 
$x$-space. Although these $e$-fluctuations have no counterpart in the
spacetime independent gauge fixing parameters of gauge theory the SLF
formalism shares with gauge theory the $e$- respectively gauge-independence
of scattering amplitudes \cite{beyond}. Off shell correlation functions are
independent of inner $e^{\prime}s$ but depend on the (fluctuating) $%
e^{\prime}s$ of the fields in the vacuum expectation values. Second order
off-shell calculations have been started in \cite{Fe-Mu}.

One expects that their leading short distance behavior is shared with that
of gauge theory but anticipates significant differences in the long distance
regime where the incorrect localization of gauge theory has its incorrect
localization properties of gauge dependent fields have their strongest
ramifications. In the massless QED limit these expected changes even affect
the particle structure (infraparticles) of the Hilbert space in a way which
has not been fully understood. From a physical viewpoint the terminology
"local gauge symmetry" is misleading since behind this more local looking
"symmetry" is the result of the presence of unphysical degrees of freedom.
The physical localization properties are only correctly described in a
positivity-respecting setting.

\begin{itemize}
\item \textbf{Massive scalar QED}
\end{itemize}

For scalar massive QED with $j_{\mu}=\varphi^{\ast}\overleftrightarrow {%
\partial_{\mu}}\varphi$ the appearance of a derivative leads to the well
known second order quadratic in $A$ contribution%
\begin{equation}
\ \delta(c-x^{\prime})A_{\mu}(x,e)\varphi^{\ast}(x)A^{\mu}(x^{\prime
},e^{\prime})\varphi(x^{\prime})+h.c   \label{sec}
\end{equation}
This comes about because the undetermined parameter $c\ $of the
Epstein-Glaser renormalization formalism which modifies the kinematic
time-ordering by adding a delta function term%
\begin{equation}
\left\langle T\partial_{\mu}\varphi\partial_{\nu}\varphi\right\rangle
=\left\langle T_{0}\partial_{\mu}\varphi\partial_{\nu}\varphi\right\rangle
+cg_{\mu\nu}\delta(x-x^{\prime})   \label{del}
\end{equation}
is fixed by the requirement (\ref{nor}) to $c=1.$ This insures that the in $%
A\ $quadratic contribution does not introduce a new counterterm parameter.
In the classical gauge theory this results from the substitution $%
\partial_{\mu }\rightarrow D_{\mu}=\partial_{\mu}-igA_{\mu}$ required by the
differential geometry of fibre bundles whereas in SLFT it is follows from
the causal localization property which leads to he independence of particles
and the S-matrix on string directions. \ We will refer to renormalization
terms whose parameters are fixed by the locality principle as \textit{%
induced interaction terms}. As in the previous case the fluctuations in the $%
e^{\prime}s$ become $e$-independent on-shell after adding all contributions
to a particular second order scattering process.

\begin{itemize}
\item \textbf{The abelian Higgs model}
\end{itemize}

In this case the $L$ in the $L,V_{\mu}\ $pair conditions depends explicitly
on the scalar sl escort $\phi$%
\begin{align}
L^{P} & =mA^{P}\cdot A^{P}H=L-\partial V  \label{vi} \\
L & =m(A\cdot AH+A\cdot\phi\overleftrightarrow{\partial}H-\frac{m_{H}^{2}}{2}%
\phi^{2}H)  \nonumber \\
V^{\mu} & =m(A^{\mu}\phi H+\frac{1}{2}\phi^{2}\overleftrightarrow {%
\partial^{\mu}}H),\text{ }Q_{\mu}=m(A^{\mu}uH+u\phi\overleftrightarrow {%
\partial_{\mu}}H)  \nonumber
\end{align}
Here the vector meson mass $m$ factor in front accounts for the correct mass
dimension $d_{sd}=$4 of the interaction density\footnote{%
Note that according to its definition the escort $\phi$ has mass dimension $%
d_{m}=0$.}. The requirement of second and third order preservation of the
pair property in the tree approximation comes with a surprise: in addition
to the expected delta contributions $\delta(x-x^{\prime})A\cdot A\phi^{2}$
and $\delta(x-x^{\prime })A\cdot AH^{2}$ which can be encoded into a $%
T_{0}\rightarrow T$ change of time-ordering, there is a second order \textit{%
induced potential} of the form of a linear combination of $%
H^{3},H^{4},\phi^{2}H^{2},\phi^{4}$ \cite{pecul} There is a formal
similarity with the gauge theoretic calculations in \cite{Scharf}.

This similarity should however not be permitted to obscure the significant
conceptual difference: whereas the induction in the SLFT setting is a direct
consequence of the perturbative implementation of the causal localization
principles, the BRST gauge theory results from the imposition of a formal
symmetry which rescues a perturbative subtheory (local observables, the
pserturbative S-matrix) from a positivity- and causal localization-
violating point-like description\footnote{%
As pointed out before the $e$-independence of the S-matrix is not a
postulate but rather the consequence of the LSZ scattering formalism in the
presence of a mass gap \cite{B-F}}.\ That SLFT contains only physical
degrees of freedom would certainly have pleased Rudolf Haag (section 5).

SSB applies to internal symmetries of interacting $s<1$ matter for which the
same field content can perfectly exist in the form of SSB or less symmetry
(independent coupling parameters and masses and a reduced number of
conserved currents). For $s\geq1$ the causal localization principle leads to
the new phenomenon of a fibre bundle like structure.

Some of these differences were suspected by Rudolf Haag and his LQP school
when they realized that their classification of superselection sectors led
to inner symmetries and the exclusion of parastatistics but confronts
serious conceptual problems in attempts to construct field algebra which
extends the local observables of gauge theory \cite{B-R} \cite{BCRV}. The
perturbative SLFT adds the new viewpoint of a causal localization caused
instrinsic quantum fibre-bundle structure in $s\geq1$ interacting theories.
The suggestion of perturbative SLFT is that interpolating fields for
particles in interacting models involving $s\geq1$ fields exist only in the
form of sl Wightman fields.

The arguments in this subsection are a good preparation for understanding
the true physical reasons why Higgs fields are needed in the presence of 
\textit{self-interacting} massive vector mesons which will be addressed
below.

\begin{itemize}
\item \textbf{Self-interacting massive vector mesons}
\end{itemize}

An even bigger surprise arises in the presence of selfinteracting massive
vector mesons. In this case the pair condition and its second order
iteration leads to two quite remarkable observations. On the one hand the
second order restriction requires the first order \ $f_{abc}$ coupling
strengths in the general ansatz for a self-interacting vector potential 
\begin{equation}
L=\sum_{abc}f_{abc}F_{a}^{\mu\nu}A_{\mu,b}A_{v,c}+..A,\phi\text{ }contr. 
\label{YM}
\end{equation}
to fulfill the Lie algebra relations of a reductive.Lie group; to find this
Lie algebra structure one has to implement the $L,Q_{\mu}$ pair condition%
\footnote{%
The $Q_{\mu}$ formalism is somewhat easier to handle and maintains a formal
similarity with the CGI gauge formulation \cite{Scharf}.} up to second
order. In the BRST gauge setting this has been known for a long time \cite%
{Scharf} but this is hardly surprising since this BRST formalism resulted
from achieving formal compatibility between Lagrangian quantization of
classical gauge theory (where this relation follows from the fibre bundle
requirements) with the algebraic structure of QFT.

But in the new sl setting the Lie algebra structure follows from the $%
s\geq1\ $causal localization principles in the form of the $L,Q_{\mu}$ pair
requirement; the calculation is in this regard formally similar to that
based on the BRST gauge formalism \cite{Scharf}.

Another important observation is that the second order leads to $d_{sd}=5\ $%
delta contribution which, if left uncompensated, would destroy the
renormalizability and hence the perturbative existence of the model. It is
saved as a renormalizable QFT by \textit{extending the field content} and
adding a nonabelian $A\cdot AH$ interaction of the massive vector mesonsn
with a $H$-field whose second order contribution contains (after adjusting
its coupling strength) compensating $d_{sd}=5$ second order terms generates
such a second order compensating\footnote{%
Contrary to its abelian counterpart for which a corresponding second order $%
d_{sd}=5$ term vanishes on the $e=e^{\prime}$ diagonal,$\ $the nonabelian
contribution provides precisely the necessary compensating contribution.}.
This is reminiscent of short distance compensations between different spin
components in supermultiplets except that in the present case it is not an
epiphenomenon of an extended symmerry but rather the raison d'\^{e}tre for
the $H$-particle.

\begin{itemize}
\item \textbf{Added comments}
\end{itemize}

The new string local quantum field theory (SLFT) shows many formal
similarities with the prior \textit{causal gauge invariance} (CGI)
reformulation of BRST in the Epstein-Glaser operator setting \cite{BDSV} .
It has the advantage of clearly distinguishing between properties which
holds only on-shell such as the BRST invariance $\mathfrak{s}S=0$ of the
S-matrix and off-shell properties as SSB

SLFT does not disqualify gauge theory, it rather shows its physical
limitations. Before commenting on this it is interesting to recall what Haag
said about the BRST formulation. In his reminiscences \cite{Haag} one finds
the following remarks "this elegant scheme is generally accepted today as
the adequate formulation of the local gauge principle in perturbation
theory. But it bears no resemblance to the conceptually simple picture in
classical theory with its continuous group acting on the fibres of a bundle."

He goes on to expresses his problem with the ghost degrees of freedom which
at the end of the day have to be removed with the help of BRST gauge
invariance. This is necessary in order to recover the most important
property of any quantum theory namely the positivity which secures the
quantum theoretical probability interpretation. Indeed the problem with pl $%
s\geq1$ interactions reveal a deep clash between pl localization and
positivity. Either one permits negative contributions in sums over
intermediate states as in the Krein space setting of local gauge theory, or
one saves positivity and uses the more natural SLF formulation in terms of $%
L,V_{\mu}$ pairs\footnote{%
The physically preferred choice is supported by the B-F theorem \cite{B-F}
which states that in the presence of a mass gap one needs no weaker
localized interpolating fields than sl fields (i.e. no need for "branes" in
LQP)}.

In philosophical terms one may say that SLFT is the result of applying
Ockham's razor to the "ghostly" BRST Krein space setting. This leads to the
concept of sl escort fields which depend on the same degrees of freedom as
those already contained in the fields which they are escorting..Such free
fields have necessarily mixed two-point functions i.e. all nondiagonal
contributions $\left\langle A^{P}\phi\right\rangle ,$ $\left\langle
A\phi\right\rangle ,$ $\left\langle A^{P}u\right\rangle ,..$of the linear sl
(Borchers) equivalence class are nonvanishing. This makes perturbative sl
calculations somewhat more involved than those in the pl Krein space
renormalization theory.

Interactions which involve $s=1$ fields are subject to additional
requirements; In the CGI gauge setting this is the BRST invariance of the
S-matrix $\mathfrak{s}S=0,$ whereas in the SLFT formulation the causal
localization requirement demands the independence of the S-matrix from the
fluctuating string directions which in $n^{th}$ order reads%
\[
d_{e}^{(n)}S^{(n)}=0,\text{ ~}d_{e}^{(n)}:=\sum_{i=1}^{n}d_{e_{i}}
\]
\ \ In the SLFT setting this requirement on the S-matrix follows directly
from the causal localization principle whereas in the CGI setting it is part
of the BRST formalism (whose spacetime interpretation is restricted to gauge
invariant observables).

For the implementation one uses the $L,Q_{\mu}\ $pair\ property and its
higher order extension. The main difference between couplings of vector
potentials to complex and Hermitian matter is that in the latter case one
obtains a richer set of second order induced terms including
selfinteractions of the $H$ and the $\phi$ escort fields. The fact that
these induced contributions have the appearance of a field-shifted Mexican
hat potential does not mean that the result bears a relation to the physics
of SSB.

A renormalized interaction density is uniquely fixed in terms of its field
content (including their masses and internal symmetries). The interpretation
of a renormalized model of QFT cannot be described by the calculating
theoretician, it is uniquely determined by intrisic properties; $Q_{sc}=0,\
Q_{sym}<\infty,\ Q_{SSB}=\infty$ represent the 3 different mutually
exclusive realizations of the causal localization principles, they
correspond to screening, inner symmetry and SSB.

As mentioned in the previous subsection selfinteracting vector mesons lead
to two new phenomena. Such models are subject to the SLFT renormalization
theory based on the $L,V_{\mu}$ pair condition. This requires the $f_{abc}$
selfcouplings to obey a fibre-bundle like structure (\ref{YM}) which, in
contrast to gauge theory, is not imposed but by quantum adjustments to
classical fibre bundles but rather a consequence of the causal localization
principles. For massive self-interacting vector-mesons there is the
additional phenomenon of second order violation renormalizability violation
which requires the compensatory presence of $H$ fields in order to save the
Standard Model.

It is interesting to note that apart from the particle-antiparticle symmetry
Nature has no use of the concept of inner symmetries and their SSB (apart
from phenomenological applications by theorists). As the success of the
Standard Model shows Nature prefers the fibre-bundle like structure of $%
s\geq1$ selfinteractions and renormalizability-saving compensations (the
raison d'\^{e}tre for the $H$).

The next and final section contains remarks about possible extensions to
higher spins $s\geq2$ and the challenge their perturbative verification
would pose to LQP.

\section{New challenges to Local Quantum Physics}

The new perturbative SLFT originated from modular localization theory within
Haag's LQP; in particular the observation that Wigner's infinite spin
representations does not permit compact localization and requires the
construction of sl fields \cite{MSY} played an important catalyzing role.
This begs the question whether the extension of perturbation theory could
also lead to an enrichment of LQP.

One challenging question is if the sl nature of interpolating fields in the
presence of $s\geq1$ massive particles is a general
(perturbation-independent) structural property of LQP. The naturalness of sl
localization\footnote{%
Weaker localization on e.g. spacelike branes is not needed (\cite{Haag}
section IV,3).} suggest that this is the case. Part of the problem of
proving such a conjecture is that one has no intrinsic nonperturbative
spacetime local (off-shell) understanding of "interaction"; the reference to
the nontriviality of the global (on-shell) S-matrix is too far removed from
properties of causal localization. The reformulation of "axiomatic" QFT in
the sense of Wightman \cite{St-Wi} in terms of sl Wightman fields is
expected to be straightforward apart from the sl replacement of the pl
extended tube analyticity (since the representation of sl fields in terms of
line integrals over pl fields is limited to free fields).

An important part of such a reformulation is a better understanding of
problems which are outside the physical range of gauge theory, as e.g. the
construction of electrical charge-carrying fields in terms of properties of
local observables \cite{Haag}. The use of the Gauss law in the LQP setting
shows that such fields are necessarily string-local in a very strong sense 
\cite{Bu}\footnote{%
As a result of photonic vacuum polarization clouds along the space-like
string direction the strings are "rigid" and, different from massive vector
mesons, cause a spontaneous symmetry breaking of the Lorentz symmetry.}. In
fact a physical description of the Hilbert space of QED (which is not a
Wigner-Fock space !) and the operators acting in it is still outstanding.

The particle structure of the Hilbert space is synonymous with the existence
of the S-matrix i.e. with the large time behavior of the charge-carrying
fields. Observationally important momentum space prescriptions for
photon-inclusive cross sections are no during replacement for a spacetime
understanding of infrared aspects. Since the gauge theoretic indefinite
metric destroys the physical localization, the correct spacetime properties
require the use of the positivity preserving sl localization; in fact the
correct analogy of quantum mechanical long range (Coulomb) interactions are
rigid (i.e. consistent with the Gauss law) string-local quantum fields.

The proposed physical spacetime explanation for the appearance of the
logarithmic divergencies is that the coupling to massless photons changes
the mass-shell delta functions of the charge-carrying massive particles into
a milder coupling strength dependent singularity which leads to vanishing
spacetime scattering amplitudes. The logarithmic divergencies are the result
of an illegitimate expansion of the "softened" mass shell singularity into a
power-series in the coupling strength accounts. In \cite{YFS} one finds
rather convincing arguments that the introduction of an infrared cutoff
parameter\footnote{%
In the SLFT formulation one would preserve covariance by viewing QED as a
massless limit of a vector mesons.} and taking the limit of its vanishing
after summing over the leading logarithms to all orders indeed leads to a
vanishing amplitudes of photonless collisions of charge-carrying particles..

The correct spacetime scattering theory is expected to be a description in
terms of a large time behavior of expectation values (probabilities). This
is outside the range of gauge theory and can only be achieved within a
positivity preserving sl setting.

Another ambitious project outside the range of gauge theory is a LQP
understanding of confinement. Different from the on-shell infrared
phenomenon whose cause is a change of the mass-shell properties of charged
particles (which leads to the vanishing of the large time limits of fields
but has no direct effect on fields and their vacuum expectation values),
confinement is a more radical phenomenon in which correlation functions
containing self-interacting massive vector mesons (massive Yang-Mills
fields) coupled to spinor or scalar quarks and the fields disappear in the
massless gluon limit and only leave their composite hadron-, gluonium- and
quark-antiquark string-bridged fields behind.

In analogy with the vanishing scattering amplitudes for photonless charged
particle collisions one expects that all correlation functions which contain
in addition to hadron and gluonium fields as well as string-bridged $q$-$%
\bar{q}$ fields also gluon or quark fields vanish, so that only those which
contain no gluon and quark operators are nontrivial. The only known way to
describe theories in which the basic model-defining fields leave only their
"composite shadows" behind in our present perturbative setting is in the
form of zero mass limits of the conceptually much clearer situation of
selfinteracting between massive vector mesons.

Do the perturbative correlation function show such a behavior? A systematic
construction of massless correlation functions of nonabelian gauge theories
can be found in \cite{Hol} and the for the present purpose relevant result
is that there are no infrared divergent correlation functions in covariant
gauges apart from the expected on-shell logarithmic divergencies which are
already present in the abelian case. This had to be expected in view of the
fact that gauge dependent fields, although possibly revealing the correct
short distance behavior in the sense of having the physically correct
beta-function\footnote{%
The Callen-Symanzik equation and in particular the beta function may turn
out to be be independent of $e.$} will be maximally incorrect for long
distances where the string-localization plays an important role.

In the presence of SLFT perturbative self-interacting gluons one however
expects such logarithmic divergences. SLFT corresponds to the noncovariant
axial gauge which has been abandoned since it generates an entangled mix of
incurable ultraviolet and infrared divergencies. But the role of $e$ in SLFT
is very different from that of a gauge fixing parameter. In contrast to a
global gauge parameter the $e$ in SLFT is as $x$ a spacetime variable in
which each field fluctuates independently in such a way that on-shell
objects as particles and the S-matrix as well as pl local observables remain 
$e$-independent, but fields and their composites depend on $e$ and transform
covariant as linear spacelike strings $\mathcal{S}=x+\mathbb{R}_{+}e,\
e^{2}=-1.$ A low order calculation of two-point correlations correlation
functions for self-interacting massless vector mesons which could reveal
whether SLFT contains a signal of confinement is more elaborate but feasible.

The SLFT renormalization theory enlarges the number of renormalizable
positivity maintaining interactions. There are two requirements which a
prescribed field content containing $s\geq1$ fields must fulfill in order to
define a positivity maintaining renormalizable SLFT. There must exist a $%
L,V_{\mu}$ pair with $d_{sd}(L)\leq4$ which fulfills the pair requirement $%
d_{e}L-\partial^{\mu}V_{\mu}=0$ and it must be possible to compensate
induced higher order anomaly terms with $d_{sd}\geq5$ by extending the field
content of $L.$

The first requirement is a lowest order consistency condition which prevents
the short-distance improving string-localization of fields to destroy the
large time field-particle relation and maintains on-shell objects as the
S-matrix $e$-independent. Its preservation in higher orders is a
normalization condition which leads to induced higher order contributions.
In contrast to renormalization counterterms which enlarge the number of
coupling parameters, higher order induced terms preserve them. They are
similar to the second order induced $A\cdot A\left\vert \varphi\right\vert
^{2}$ term in scalar QED except that in SLFT they do not originate from
quantization of classical fibre bundle structures but are an autonomous
consequence of the positivity maintaining causal localization principle of
QFT. This applies also to the Lie-algebra structure of self-interacting
vector potentials. It shows that QFT does not need "quantization crutches"
but can perfectly stand on its own feet.

The second requirement maintains renormalizability to all orders. It has no
counterpart in $s<1$ pl interactions for which first order renormalizability 
$d_{sd}(L)\leq4$ guaranties renormalizability to all orders (no
"induction"). It is a new phenomenon for interactions involving $s\geq1$
fields (unless one wants to view it as an analog of the alleged
renormalizability-improving role of compensation between different spin
components within a supermultiplet).

Interactions of abelian vector mesons with spinor-, complex scalar- or real
(Higgs)- matter do not require the compensatory extension of the field
content; the implementation of the pair condition suffices in those models.
The need for a compensatory enlargement in order to preserve second order
renormalizability leads to \textit{the Higgs field in the presence of
massive self-interacting vector potentials} \cite{beyond} \cite{pecul}. Both
requirements have their counterpart in the CGI operator setting of BRST
gauge theory \cite{Scharf} where the causality implementing pair requirement
corresponds to the BRST invariance of the S-matrix.

The SLFT renormalization theory is still in its infancy. For $s>1$ there are
as yet no SLFT results apart from the qualitative observation that higher
spin fields will enhance the short distance dimension of $Q_{\mu}$ which in
turn may lead to renormalizability violating induced higher order delta
terms whose compensation requires the enlargement of the field content. The
most plausible scenario in analogy to the compensatory role of the Higgs
field is that the highest spin $s$ requires the presence of all lower spin
fields (e.g. for $s=2$ the presence of $s=1$ and $s=0$)

Fields belonging to the zero mass infinite spin Wigner class fail on the $%
L,Q_{\mu}\ $pair requirement; they exist only in the form of sl free fields 
\cite{dark}. We will refer to such matter as \textit{non-reactive or inert}
in the sense of SLFT perturbation theory. Hence the problem posed by the two
requirements is the question: \textit{up to what spin does matter remain
reactive }?

Since a further lessening of the tightness of localization beyond sl as a
localization on spacelike hypersurfaces ("branes") brings no gain for
renormalizability, it is not unreasonable to expect that a field content
which permits no renormalizable perturbative interaction in the SLFT
formulation has also no counterpart outside perturbation theory. This belief
is based on the naturalness of string-localization i.e. the fact that
particles in LQP always admit sl interpolating fields with pl being a
special case of sl \cite{B-F}.

This does not require the convergence of the perturbative series; the
singular nature of fields due to the omnipresence of vacuum polarization
clouds limits their use in mathematical existence proofs. But a field of
spin $s>1~$which allows no renormalizable interactions with itself and lower
spin fields is also believed to be inert par excellence.

All positive energy matter can be shown to admit a conserved energy-momentum
tensor; the No-Go theorem in \cite{WW} for \textit{massless} higher spin
matter refers to pl fields, whereas conserved weaker localized sl E-M
tensors whose global charges are identical to those of their pl siblings
exist and have a well-defined massless limit. Hence also inert matter which
only exists in the form of free fields couples to gravity and leads to
gravitational backreaction, which makes it interesting as candidates for
dark matter. Intrinsically sl infinite spin matter is inert \cite{dark} but
as a result of its fleeting nature resulting from its masslessness it does
not seem to be compatible with the halo like accumulation of dark matter
around galaxies.

This is the content of a structural theorem of LQP which states that in
order to describe particles one does not need weaker than string localized
interpolating fields. Hence one expects that interactions which fail on both
previous properties do not exist as the result of lack of reactivity of the
highest spin component.

The string-localization of matter fields in interactions involving sl $s\geq1
$ potentials in SLFT renormalized perturbation theory begs the question to
what extend its occurrence can be understood in the nonperturbative LQP
setting. The problem is that there is no nonperturbative localization-based
intrinsic definition of "interaction"; the existence of a nontrivial
S-matrix is too remote from the spacetime properties of interacting fields.
A proof would amount to a theorem stating that a particle spectrum with mass
gaps which includes $s\geq1$ particles is either a free field theory or an
model whose interacting fields are sl Wightman fields. The adjustment of
Wightman's axiomatic framework to sl fields would then be the lesser problem.

Perturbative SLFT also directs attention to a new problem of formal
symmetries which are not inner symmetries in the sense of the DHR
superselection theory. As mentioned before such a problem is posed by the
perturbative Lie algebra structure of self-interacting vector mesons. Such a
situation can not be subsumed under inner symmetry since the latter always
permit interactions with the same field content but less or no symmetry.
There is as yet no natural conceptual place in LQP.

In the BRST gauge formulation this is less surprising since that formalism
is the result of a repair job which is necessary to control the indefinite
metric aspects of a formulation obtained from adjusting a classical fibre
bundle setting to the exigencies of a quantum theory which is only possible
at the price of indefinite metric and ghosts. In a somewhat metaphoric sense
result from the quantization (of a classical theory which has no use for
"positivity"). Why does Nature not present particle multiplets associated to
internal symmetries (or Goldstone particles of an exact SSB)? Why does she
prefers the Lie algebra structure of self-interacting vector mesons (the
Standard Model) ?

LQP is still far from its ultimate goal of establishing the mathematical
existence of nontrivial models and finding mathematically controlled
approximation procedures. However there are good reasons to expect that the
pursuit of this goal will lead to important more new insights. Rudolf Haag's
general LQP view of QFT as causally localized quantum matter \cite{rem}
remains a valuable compass which helps to avoid a cul-de-sac as that
mentioned in section 5.

\begin{acknowledgement}
The last two sections are part of an ongoing joint project with Jens Mund.
Its ultimate aim is to replace local gauge theory by a formulation which is
compatible with Haag's LQP. For a critical reading I am indebted to Joe
Varilly.
\end{acknowledgement}

\end{document}